\documentclass[pre,twocolumn,floatfix,tightenlines,nofootinbib,superscriptaddress,showpacs]{revtex4}%

\usepackage{amsmath}
\usepackage{amssymb,amsfonts}
\usepackage{dcolumn}
\usepackage{bm}
\usepackage[mathcal]{euscript}
\usepackage[pdftex]{graphicx}
\graphicspath{{figs/}}
\usepackage{pdfpages}
\usepackage{epsfig}
\usepackage{subfigure}
\usepackage{color}
\usepackage{url}

\definecolor{darkblue}{rgb}{0,0.08,0.45}
\newcommand{\ie}{\textit{i.e.}}

\newcommand{\mathnotation}[2]{\newcommand{#1}{\ensuremath{#2}}}

\renewcommand{\l}{\left}			
\renewcommand{\r}{\right}			
\mathnotation{\xb}{\mathbf{x}}
\mathnotation{\vb}{\mathbf{v}}
\mathnotation{\htop}{h} 
\mathnotation{\htopb}{\htop_{\mathrm{braid}}}   
\mathnotation{\htopf}{\htop_{\mathrm{flow}}}    
\mathnotation{\id}{\mathrm{identity}}           
\mathnotation{\e}{\mathrm{e}}
\mathnotation{\ldef}{\mathrel{\raisebox{.069ex}{:}\!\!=}}
\mathnotation{\rdef}{\mathrel{=\!\!\raisebox{.069ex}{:}}}
\mathnotation{\bi}{b}				
\mathnotation{\wb}{w_{\mathrm{B}}} 
\mathnotation{\lm}{\lambda_{M}}
\mathnotation{\map}{\mathcal{M}}
\mathnotation{\tw}{\tau_{\mathrm{W}}}  
\def\d{{\mathrm d}}
\mathnotation{\tnot}{t_0}
\mathnotation{\tinf}{t_\mathrm{inf}}
\mathnotation{\vpara}{v_\parallel}
\mathnotation{\vperp}{v_\perp}
\mathnotation{\xpara}{x_\parallel}
\mathnotation{\xperp}{x_\perp}
\mathnotation{\xf}{x_{\mathrm{f}}}
\mathnotation{\T}{T} 

\newcommand{\degree}{\ensuremath{^\circ}}
\mathnotation{\BM}{f} 
\mathnotation{\g}{g} 
\mathnotation{\lw}{\ell_{\mathrm{W}}}
\mathnotation{\lb}{\ell_{\mathrm{black}}}
\mathnotation{\lac}{\Lambda} 
\mathnotation{\A}{a}
\mathnotation{\Aa}{A}
\mathnotation{\reight}{r_\infty}
\mathnotation{\dotd}{\skew{8}\dot{d}}
\mathnotation{\stripw}{\Delta} 
\mathnotation{\ti}{\tau_{\mathrm{i}}} 
\mathnotation{\PC}{P(C)} 
\mathnotation{\PT}{\tilde{P}(t)} 
\mathnotation{\Cg}{C_\mathrm{g}}
\mathnotation{\mueigen}{\mu}

\begin{document}


\title{Slow decay of concentration variance due to no-slip walls in
  chaotic mixing}

\author{E. Gouillart}
\affiliation{Surface du Verre et Interfaces, UMR 125 CNRS/Saint-Gobain, 93303
Aubervilliers, France}
\affiliation{Service de Physique de l'Etat Condens\'e, DSM, CEA Saclay,
URA2464, 91191 Gif-sur-Yvette Cedex, France}
\author{O. Dauchot}
\author{B. Dubrulle}
\affiliation{Service de Physique de l'Etat Condens\'e, DSM, CEA Saclay,
URA2464, 91191 Gif-sur-Yvette Cedex, France}
\author{S. Roux}
\affiliation{LMT-Cachan, UMR CNRS 8535/ENS-Cachan/Univ. Paris
VI/PRES UniverSud , 94 235 Cachan, France}
\author{J.-L. Thiffeault}
\affiliation{Department of Mathematics, University of Wisconsin,
  Madison, WI 53706, USA}

\date{\today}


\begin{abstract}

  Chaotic mixing in a closed vessel is studied experimentally and
  numerically in different 2-D flow configurations. For a purely
  hyperbolic phase space, it is well-known that concentration
  fluctuations converge to an eigenmode of the advection-diffusion
  operator and decay exponentially with time.  We illustrate how the
  unstable manifold of hyperbolic periodic points dominates the
  resulting persistent pattern. We show for different physical viscous
  flows that, in the case of a fully chaotic Poincar\'e section,
  parabolic periodic points at the walls lead to slower (algebraic)
  decay. A persistent pattern, the backbone of which is the unstable
  manifold of parabolic points, can be observed.  However, slow
  stretching at the wall forbids the rapid propagation of stretched
  filaments throughout the whole domain, and hence delays the
  formation of an eigenmode until it is no longer experimentally
  observable. Inspired by the baker's map, we introduce a 1-D model
  with a parabolic point that gives a good account of the slow decay
  observed in experiments. We derive a universal decay law for such
  systems parametrized by the rate at which a particle approaches the
  no-slip wall.

\end{abstract}
\vspace{-0.5cm}

\maketitle

\section{Introduction}
\label{intro}

Many industrial applications involve the mixing of viscous fluids.
Fields as diverse as chemical engineering, the pharmaceutical and
cosmetics industries, and food processing depend on the stirring of
initially heterogeneous substances to obtain a product with a
sufficient degree of homogeneity.  Viscous, confined or fragile fluids
are best mixed by non-turbulent flows, which tend to be less effective
at mixing than their turbulent counterpart.  However, some laminar
flows exhibit \emph{chaotic advection}, meaning that they have chaotic
Lagrangian trajectories~\cite{Aref1984, Ottino1989}, allowing them to
rival turbulent flows in their ability to mix. The framework of
chaotic advection and dynamical systems provides a useful
characterization of mixers that relies on the nature of the phase
space, or the stretching statistics of Lagrangian trajectories.
However, an essential issue for the mixing of a diffusive passive
scalar is to predict the rate at which the scalar concentration is
homogenized by a given stirring protocol.

Various approaches including an eigenmode
analysis~\cite{Pierrehumbert1994, Fereday2002, Wonhas2002,
  Sukhatme2002, Pikovsky2003, Thiffeault2003, Fereday2004,
  Thiffeault2004b, Liu2004, Haynes2005}, a large-deviation description
of the stretching distribution~\cite{Antonsen1995,Antonsen1996}, and
multifractal formalism~\cite{Ott1989,Antonsen1991} have provided
insights into the structure of the mixing pattern and its decay rate.
Of particular importance in some of these studies is the idea that for
time-periodic flows the spatial mixing pattern becomes
\emph{persistent}, in the sense that it repeats itself in time but
with a decreasing overall amplitude of fluctuations.  Time-persistent
spatial patterns have been observed in numerical
simulations~\cite{Pierrehumbert1994, Sukhatme2002, Fereday2004}, as
well as in dye homogenization experiments in cellular
flows~\cite{Rothstein1999,Jullien2000, Voth2003}, and have been
related to the slowest decaying eigenmode of the advection-diffusion
operator.  The term \emph{strange eigenmode}, originally coined by
Pierrehumbert~\cite{Pierrehumbert1994}, is used to describe these
patterns.  The eigenmode amplitude decays exponentially with time at
a rate determined by its associated eigenvalue, and an exponential
decay of concentration variance has indeed been observed in various
systems \cite{Pierrehumbert1994, Rothstein1999,Jullien2000,
  Sukhatme2002, Voth2003, Thiffeault2003, Fereday2004, Haynes2005}.

However, such results were obtained either in idealized systems
\cite{Pierrehumbert1994, Fereday2002, Wonhas2002, Sukhatme2002,
  Thiffeault2003, Fereday2004, Gilbert2006}, or in cellular
flows~\cite{Rothstein1999, Jullien2000, Voth2003}. It has been
suggested~\cite{Chertkov2003a,Lebedev2004, Schekochihin2004,
  Salman2007} that mixing might be slower in large-scale bounded
flows, because of slow stretching dynamics in the vicinity of a
no-slip wall. The specific form of the velocity field at a no-slip
wall was first noticed and exploited by Chertkov and
Lebedev~\cite{Chertkov2003a}. The authors calculated concentration
statistics by ensemble averaging over different realizations of a flow
in a bounded domain with random time-dependence. They obtained a
transient algebraic decay of the scalar variance attributed to the
influence of the wall, followed by an asymptotic exponential phase.
Shortly thereafter, experiments of elastic turbulence in a
microchannel~\cite{Burghelea2004} showed an anomalous scaling of
mixing dynamics with the P\'eclet number, which the authors related to
the predictions of Chertkov and Lebedev~\cite{Chertkov2003a}. Detailed
numerical simulations of scalar advection by a short-correlated flow
in a bounded domain were recently performed by Salman and
Haynes~\cite{Salman2007}, who characterized the scalar decay with a
multi-stage scenario that includes a transient algebraic decay. All
theoretical and numerical studies~\cite{Chertkov2003a,Lebedev2004,
  Salman2007} assumed that, in bounded flows, scalar fluctuations are
rapidly completely exhausted in the bulk because of efficient
stretching therein, while scalar inhomogeneity subsists only in a
decreasing pool at the boundary.

In a previous letter~\cite{Gouillart2007}, we have reported on the
first experimental observation of ``slow'' algebraic mixing dynamics
imposed by a no-slip wall in a deterministic \hbox{2-D} chaotic
advection protocol.  In the present paper, we explore in more detail
the successive stages of mixing of a passive scalar in experiments and
simulations of Stokes flow.  For several chaotic advection protocols,
a blob of dye is released in a closed vessel and homogenized.  In
contrast with all studies mentioned above, we focus here on the
influence of the wall on the concentration field far from the
boundary, which we show is contaminated by the algebraic dynamics near
the wall. Our approach is based on a Lagrangian description of
stretched filaments slowly fed from the wall into the bulk. A
simplified one-dimensional model --- a generalization of the baker's
map --- allows us to describe the various mechanisms at play and to
reproduce the main features of the evolution of the concentration
probability distribution.

The paper is organized as follows. In Section~\ref{sec:mechanisms} we
review the main ingredients of chaotic mixing.  We discuss the successive
stages of mixing and the associated length scales, which are then
illustrated on the pedagogical example of the well-studied hyperbolic
baker's map.  For this ideal system, we relate the structure of the
strange eigenmode to the unstable manifold of the least-unstable periodic
point.  Section~\ref{parabolic} is the core of the paper.  We first
report on homogenization experiments conducted with a figure-eight
protocol already described in~\cite{Gouillart2007}, which we complement
by numerical simulations of a viscous version of the blinking vortex
flow~\cite{Aref1984}, and a modified version of the baker's map with a
parabolic point at the wall. In all cases, we observe anomalously slow
mixing, that is an algebraic --- rather than exponential --- decay of
concentration variance.  We argue that this behavior is generic for
two-dimensional mixers where the chaotic region extends to fixed no-slip
walls. In such systems, poorly stretched fluid escapes the wall at a slow
rate (controlled by no-slip hydrodynamics) through the unstable manifold
of parabolic points on the wall. These poorly-mixed blobs contaminate the
whole mixing pattern, up to the core of the domain where stretching is
larger. We show that the modified baker's map describes the experiments
qualitatively and allows an analytic derivation of the observed scalings
for the concentration distribution.  A discussion on very long times,
general initial conditions and hydrodynamical optimization is finally
presented in Section~\ref{sec:discuss}.

\section{Homogenization mechanisms}
\label{sec:mechanisms}

\subsection{Stages and length scales of mixing}

\mathnotation{\lblob}{\ell_{\mathrm{blob}}}
\mathnotation{\lv}{\ell_{\mathrm{v}}}

\begin{figure}[t!]
\centerline{\includegraphics[width=0.9\columnwidth]{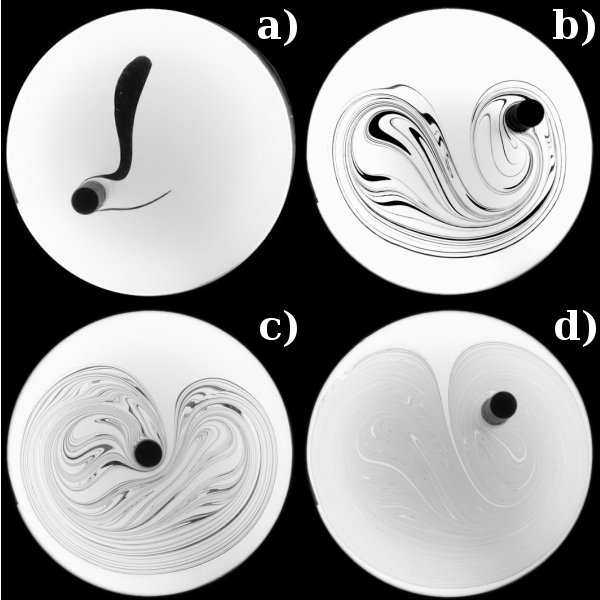}}
\caption{Successive stages of homogenization for a blob of dye stirred
  by the figure-eight protocol (see section~\ref{parabolic} for
  details).  (a) An initial blob is stretched by gradients in the
  velocity field. At early times, stretched filaments are still
  too broad for diffusion to be noticeable, and the concentration
  variance is constant. (b) As time increases, filaments are stretched and
  folded repeatedly, while a strip of white
  fluid coming from the boundary is inserted periodically at the
  core of the mixing pattern. As a result of this mass injection, the
  filamentary pattern grows slowly towards the boundary with time,
  while filaments become thin enough (c) for diffusion to become
  effective and cause the strips of dye to become more gray.  Different gray
  levels correspond to different stretching histories.  (d) Later,
  different filaments start interpenetrating, and the
  concentration field results from the averaging of concentration
  values coming from neighboring strips.}
\label{figs_closedhom:homsteps}
\end{figure}

In this section, we describe briefly how the concentration field of a
passive scalar (e.g. dye) evolves from an initially segregated state
towards homogeneity. For illustrative purposes, we will consider the
example shown in Figure~\ref{figs_closedhom:homsteps} of a blob of dye
of initial scale $\lblob$ smaller than the velocity field scale $\lv$,
which is of the same order as the domain size $L$.  Very viscous
fluids typically support only laminar flows. Such flows may still
lead to complicated, that is chaotic, Lagrangian
trajectories~\cite{Aref1984,Ottino1989}.

Three different stages of the mixing process are apparent in
Figure~\ref{figs_closedhom:homsteps}. An initial blob
(Fig.~\ref{figs_closedhom:homsteps}(a)) is deformed by the stirring
velocity field.  At early times, the concentration pattern evolves as
finer scales are created, yet the variance (the spatially-integrated
squared fluctuations from the mean) is almost unchanged as the spatial
scales are still too large for diffusion to be efficient (first stage,
Fig.~\ref{figs_closedhom:homsteps}(b)).  After several stretching and
folding events, the width of a filament of dye stretched at a typical
rate $\lambda$ stabilizes at the so-called Batchelor length
\begin{equation}
  \wb\ldef\sqrt{\kappa/\lambda},
\end{equation}
with $\kappa$ the diffusivity, where the effects of compression and
diffusion balance.  Obviously, in a realistic flow, stretching is not
constant, but the width of a filament quickly adapts to the local
stretching rate $\lambda(\xb)$.  The length scale $\wb$ is the smallest
that can be observed inside the concentration pattern: an initial blob
with a scale greater than $\wb$ is stretched and folded into many
filaments that are compressed up to the diffusive scale $\wb$. During
a second stage (Fig.~\ref{figs_closedhom:homsteps}(c)), after a strip
has stabilized at the width $\wb$, the amplitude of the concentration
profile decreases according to the stretching experienced by the
strip, to ensure conservation of
dye~\cite{Balkovsky1999,ThiffeaultAosta2004}. Different gray levels
correspond to different stretching histories along the elongated image
of the initial blob. Finally, since filaments are stretched but also
folded, they are eventually pressed against each other, and their
diffusive boundaries interpenetrate (third stage,
Fig.~\ref{figs_closedhom:homsteps}(d)).  Ultimately, homogenization
takes place inside a ``box'' of size $\wb$ through the averaging of
many strips that have experienced different stretching histories and
have therefore different amplitudes~\cite{Villermaux2003}.

So far, the most satisfying explanation for the decay of inhomogeneity
in chaotic mixing has been the \emph{strange eigenmode} theory,
initially proposed by Pierrehumbert in a 1994
paper~\cite{Pierrehumbert1994}. The strange eigenmode is the second
slowest decaying eigenmode of the advection-diffusion operator (the
first trivial mode corresponds to a nondecaying uniform
concentration). We consider in the following the case of periodic
velocity fields, where the periodically-strobed strange eigenmode is
an eigenvector of the time-independent Floquet operator.  It decays at
an exponential rate fixed by the real part of the corresponding
eigenvalue.  The projection of the initial concentration field on this
eigenmode decays slower than the contributions from other eigenmodes,
so that one expects the concentration field to converge rapidly to a
permanent spatial pattern determined by the strange eigenmode, whose
contrast decays exponentially.

A simple physical motivation for the strange eigenmode is as follows. The
asymptotic concentration is governed by filaments that are pressed
against each other in a box of width~$\wb$.  But these filaments have
explored the whole domain, and hence may possess different stretching
histories~\cite{Pierrehumbert2000}.  The decay rate can thus depend on
\emph{global} properties of the flow~\cite{Fereday2004,Haynes2005}.  In
particular, it is sensitive to spatial correlations, and cannot be
expressed simply in terms of the stretching statistics.

Since the seminal paper of Pierrehumbert, strange eigenmodes have been
observed in many numerical
studies~\cite{Fereday2002,Wonhas2002,Sukhatme2002,Pikovsky2003,Fereday2004}.
The proposed evidence for strange eigenmode were (i) the onset of
permanent spatial concentration patterns and (ii) an exponential decay
for the concentration variance, whose rate depended only weakly on the
diffusion. Recurrent spatial patterns have also been observed in
experiments where a viscous fluid is stirred by an array of
magnets~\cite{Rothstein1999,Jullien2000,Voth2003}, yet the concentration
decay seemed slower than exponential. In the following section, we
briefly show how a strange eigenmode arises in a one-dimensional baker's
map, and relate the spatial structure of the eigenmode to the regions of
lowest stretching.

\subsection{Tracing out the strange eigenmode in a uniformly
  hyperbolic model}
\label{sec:SEhyp}

In this section, we describe how periodic points of a uniformly
hyperbolic map affect the concentration field $C(x,t)$ of a
low-diffusivity scalar, insofar as they determine the spatial
structure of the observed strange eigenmode. We show that the
concentration pattern obtained from an initial blob after successive
iterations of the map is determined by the least unstable periodic
point of the map, and its multifractal unstable manifold. For pedagogical
reasons, we use one
of the most-studied model of chaotic mixing, the inhomogeneous
area-preserving baker's
map~\cite{Farmer1983,Ott1989,Antonsen1991,Fereday2002}.

The area-preserving baker's map is
defined on a two-dimensional square region by dividing the region in two
strips, stretching, and re-stacking them.  It has the property of mapping
a~$y$-independent distribution to another such
distribution~\cite{Fereday2002}.  We thus take our initial blob to be a strip
uniform in the~$y$-direction, and the baker's map stretches and folds this
strip to create more strips, leaving the concentration independent of~$y$.  We
can thus focus on one-dimensional distributions that depend only on the~$x$
coordinate: they represent a `cut' across a striated pattern of strips
like the pattern in Fig.~\ref{figs_closedhom:homsteps}.  Hence,
we limit ourselves to a one-dimensional version of the baker's map which
captures the essence of dynamics.

\begin{figure}
\centerline{\includegraphics[width=0.7\columnwidth]{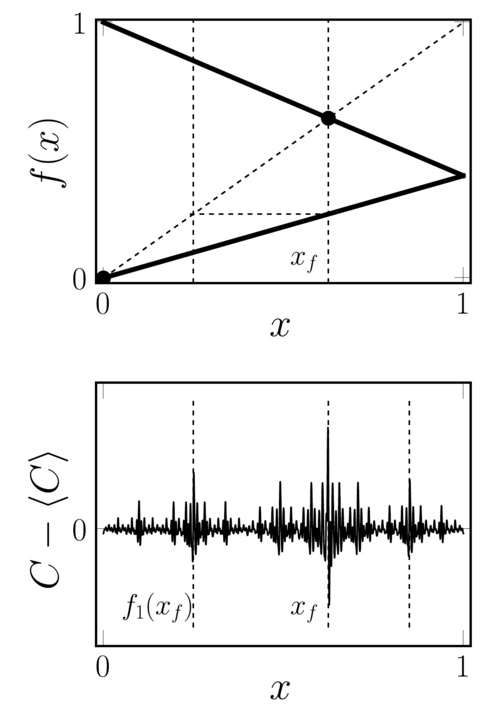}}
\caption{(a) One-dimensional baker's map $\BM$, with two fixed points at
$x=0$ and $x=1/(2-\gamma)$. For $\gamma<1/2$ the most stable fixed point is
$\xf=1/(2-\gamma)$. (b) Concentration profile obtained for an initial blob
transformed by 17 iterations of $\BM$ ($\gamma=0.4,\,\kappa=10^{-5}$). Dominant
``spikes'' are located at $\xf$ and (with decreasing amplitude)
around its iterates (of decreasing stability).\label{figs_dyna:eigen}}

\end{figure}

The baker's map~$\BM$ reads
\begin{subequations}
\begin{equation}
\BM: x \mapsto \BM_1(x) \cup \BM_2(x)
\label{eq:map_baker_a}
\end{equation}
where
\begin{equation}
\BM_1(x) = \gamma x\,;\qquad \BM_2(x) = 1 -(1-\gamma)x
\end{equation}
\label{eq:map_baker}%
\end{subequations}%
and the union ($\cup$) symbol in~\eqref{eq:map_baker_a} means that~$\BM$ is
one-to-two: every point~$x$ has two images given by~$\BM_1(x)$ and~$\BM_2(x)$.
The parameter~$\gamma$ satisfies~$0<\gamma<1$ and controls the homogeneity of
stretching, with~$\gamma=1/2$ being the perfectly homogeneous case. 
The baker's map is represented in Fig.~\ref{figs_dyna:eigen}.

Under the action of the baker's map, the concentration profile evolves as
\begin{equation}
C(x,t+1)=C\bigl(\BM^{-1}(x),t\bigl).
\label{eq:Cevolve}
\end{equation}
$\BM$ therefore
transforms the concentration profile $C(x,t)$ at time $t$ into two images
``compressed'' by respective factors $\gamma$ and $1-\gamma$. 

\begin{figure}
\centerline{\includegraphics[width=0.99\columnwidth]{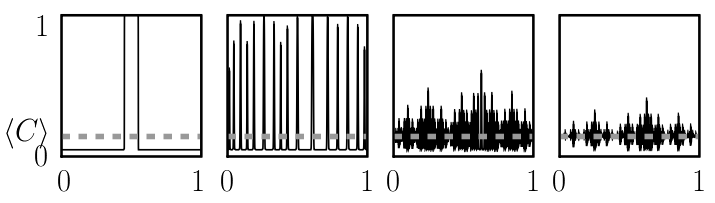}}
\caption{Main stages in the evolution of the concentration profile, from
left to right. An
initial blob is stretched into many filaments by the map. Once filaments
reach the diffusive scale $\wb$, intermediate concentration levels appear. The
concentration profile takes the form of a persistent pattern --- the
strange eigenmode --- when all boxes of size $\wb$ contain an image of
the initial unit interval.\label{figs:scenar_baker}}
\end{figure}

First, we compute numerically the evolution of an initial blob under the
action of $f$.  The initial concentration
\begin{equation}
  C(x,0) = C_0(x) = \begin{cases}
    1, \quad &x_a \leq x \leq x_b; \\
    0, &\text{otherwise,}
  \end{cases}
  \label{eq:C0}
\end{equation}
is a strip of constant concentration between~$x_a$ and~$x_b$.  Diffusion is
mimicked by letting the concentration evolve diffusively during a unit
time interval~\cite{Pierrehumbert2000,Fereday2002}.  During that interval, $C$
evolves according to the heat equation with diffusivity $\kappa$. We use
periodic boundary conditions.

Figure~\ref{figs_dyna:eigen} shows the concentration profile for a
typical simulation with $\gamma=0.4$, after $17$ iterations of $\BM$
alternated with diffusive steps.  We see that the system is well
mixed, insofar as fluctuations of $C$ around its spatial mean $\langle
C \rangle$ (which is conserved by the map) are very weak compared to
the initial blob.  Angle brackets denote a spatial average.  A closer
inspection reveals that fluctuations of $C$ have more important values
at some points, so that the concentration pattern has distinctive
spikes. Remarkably, the spatial pattern visible in
Fig.~\ref{figs_dyna:eigen}(b) is permanent, as can be seen on the two
rightmost profiles in Fig.~\ref{figs:scenar_baker}: further iteration
of the map does not change the form of $C$, only its amplitude.
This shows that the concentration field converges very rapidly to an
eigenfunction of the advection-diffusion operator, dubbed the strange
eigenmode. We measure an exponential decay of the concentration variance,
consistent with convergence to an eigenmode. Strange eigenmodes in
the baker's map, and their decay rate, have been studied in detail
in Refs.~\cite{Fereday2002,Wonhas2002,Gilbert2006}.  Here we provide a simple
and novel way to characterize the spatial structure of the strange eigenmode.
More explicitly, we describe below how the strange eigenmode pattern traces
out the unstable manifold of the periodic point of the map with weakest
stretching.  This echoes the description of invariant sets in open
flows in terms of the \emph{chaotic saddle}~\cite{Tel2000}.

As can be seen in
Fig.~\ref{figs_dyna:eigen}, the map $\BM$ has two period-1 (fixed) points, one
at~$x=0$ and another at $x=\xf \ldef 1/(2-\gamma)$.  Of course, since
the 1-D baker's map is one-to-two, both fixed points map to an
additional point, so that the fixed points have iterates other than
themselves.  For $\gamma<1/2$ ($\gamma=0.4$ in
Fig.~\ref{figs_dyna:eigen}), the second fixed point is less unstable
than the first one, as the compression factor at $\xf$ is smaller. We
notice in Fig.~\ref{figs_dyna:eigen} that the highest spike in the
concentration pattern is located at $x=\xf$, whereas spikes of
decreasing height are located at iterates of $\xf$: $f_1(\xf)$,
$f_2(f_1(\xf)),\ldots$: the unstable manifold of $\xf$ forms the
backbone of the concentration pattern.

Let us examine more closely the iteration of the concentration pattern.  After
$t$ iterations of $f$, the initial blob has been transformed into $2^t$ strips
compressed by factors
$\gamma^t,\,\gamma^{t-1}(1-\gamma),\dots,\,(1-\gamma)^t$. However, diffusion
imposes that the width of an elementary strip saturates at the Batchelor width
$\wb$ where diffusion balances stretching. Using the range of possible stretchings in the map, we have
\[
\sqrt{\frac{\kappa}{1-\gamma^2}}\leq \wb \leq
\sqrt{\frac{\kappa}{1-(1-\gamma)^2}},
\]
where the diffusivity $\kappa$ has been rescaled by the size of the
domain $L=1$ and the time-period $\T=1$.
We therefore approximate $\wb\simeq\sqrt{\kappa/(1-\Gamma^2)}$, where
\[
\log \Gamma=\gamma\log(\gamma)+(1-\gamma)\log(1-\gamma)
\]
is the Lyapunov exponent. The concentration profile of
Fig.~\ref{figs_dyna:eigen} has a typical variation scale of $\wb$.
Under repeated compression and diffusion steps, each elementary strip
converges to a Gaussian peak of width $\wb$ (see the second picture
from left in Fig.~\ref{figs:scenar_baker}), centered on iterates of
the initial blob centroid $x_c \ldef (x_a+x_b)/2$. The amplitude of
each Gaussian strip is proportional to~$\lac/\wb$ to conserve total
concentration, where $\lac$ is the multiplicative compression
experienced by the strip. \label{lambda_defined}

The strange eigenmode regime is reached when the initial blob has been
stretched enough so that its centroid has an iterate in every box of
size $\wb$ -- that is, each box contains at least one image of the initial
domain.  In this regime, the concentration $C(x,t)$ measured at a
point $x$ results from the addition of slightly shifted strips, whose
centers are all iterates of $x_c$ that fit into a ``box'' of size
$\wb$ centered on $x$. The random averaging of such strips has been
proposed as the mechanism controlling the homogenization rate
\cite{Villermaux2003,Venaille2007}. However, due to strong
time-correlations of stretching, this averaging is not a random
uncorrelated process here.  Fluctuations of $C(x)$ around $\langle C
\rangle$ are typically inversely proportional to the number of
iterates in the box centered on $x$~\cite{Villermaux2003}. High spikes
in the pattern therefore correspond to ``boxes'' with relatively few
contributing iterates --- i.e. images of the initial domain that have
experienced relatively low compression.

Iterates of the initial profile located around $\xf$ have experienced
successive $1-\gamma$ factors during the last steps of the process while
converging to the attracting periodic point $\xf$, since they have been
transformed by the second branch $\BM_2$ during all recent iterations. On
average, these iterates have experienced compressions smaller than the
mean compression $\Gamma^t$. This explains why the sharpest fluctuations
are visible around $\xf$, and with decreasing amplitude around iterates
of $\xf$ of decreasing stability. For example, all iterates around
$f_1(\xf)$ have experienced a large number of successive $1-\gamma$
compressions, and an additional $\gamma$ compression during the latest
iteration. In contrast, iterates around $\xf$ have experienced only
$(1-\gamma)$ compression steps in the last iterations, and fluctuations
are greater.

We conclude that regions of low stretching control the structure of
the concentration pattern. This behavior has already been illustrated
for the case of mixed phase space (with elliptical islands or weakly
connected chaotic domains)~\cite{Pikovsky2003,Popovych2007}, but also
holds for purely hyperbolic domains with uneven stretching.

In addition, our description of the coherent structure of the strange
eigenmode sheds a new light on why it is so difficult to predict the
decay rate of the
eigenmode~\cite{Wonhas2002,Thiffeault2003,Thiffeault2004b,
  Gilbert2006}.  Fluctuations decrease at a rate determined by the
subtle interplay of peaks corresponding to the successive iterates of
$\xf$, therefore spatial correlations of stretching histories play an
important role in the decay rate, which cannot be related easily to
the distribution of stretching in the map.  We have provided here an
example of concentration patterns dominated by the periodic
structures with least stretching.

We now turn to the experimental study of mixing by fully chaotic flows
in bounded domains. In such mixers, the dominant periodic structures
are parabolic points on a no-slip boundary, and we show that such
points impose slower algebraic dynamics.

\section{Parabolic points at the walls}
\label{parabolic}

In this section, we report on dye homogenization experiments
conducted in a closed vessel where a single rod stirs fluid with a
figure-eight motion. In this physical system, the phase portrait is not
purely hyperbolic as it was in the baker's map: we describe how
separatrices (parabolic
points) appear on the wall as a consequence of no-slip hydrodynamics.  We
show that these regions of low stretching slow down mixing and
contaminate the whole mixing pattern up to its core, far from the wall.
These experimental results were briefly presented
in~\cite{Gouillart2007}. Here, results from a numerical simulation of a
counter-rotating viscous blinking vortex
protocol~\cite{Aref1984,Jana1994} are also presented. The dye pattern
bears a strong resemblance to that of the figure-eight protocol, and we
show that parabolic points at the walls are again responsible for
algebraic decay. Inspired by the baker's map studied in
Sec.~\ref{sec:SEhyp}, we introduce a
simplified 1-D model that produces comparable algebraic mixing dynamics
for this broad class of mixers.

\subsection{Algebraic decay in experiments and numerical simulations}
\label{sec:algebraic}

\begin{figure}[t!]
\subfigure[]{
\includegraphics[width=4cm]{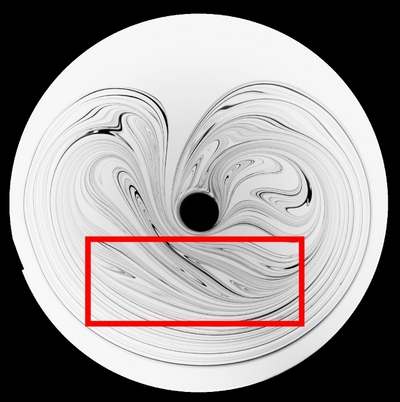}
}
\subfigure[]{
\includegraphics[width=4cm]{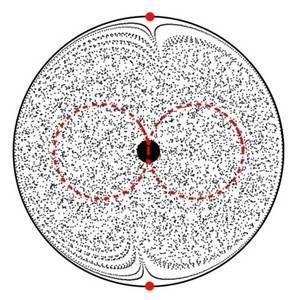}
}
\subfigure[]{
\includegraphics[width=4cm]{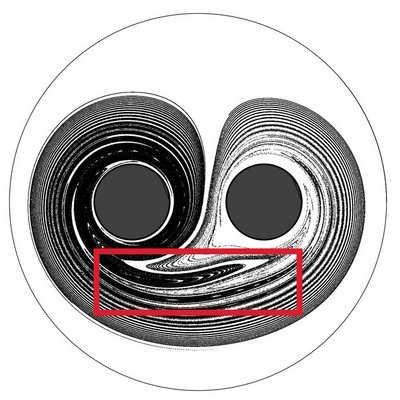}
}
\subfigure[]{
\includegraphics[width=4cm]{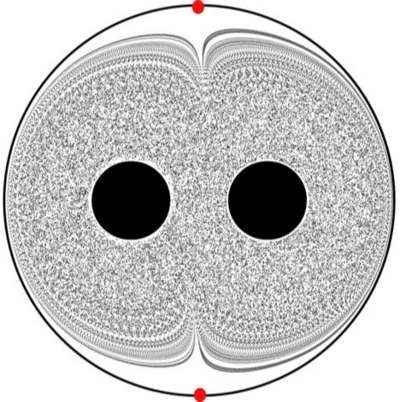}
}
\caption{[Color online] Homogenization in closed flows for the experimental
  realization of the figure-eight protocol (a)--(b) and the numerical
  simulation of the contra-rotating blinking vortex (c)--(d). The
  heart-shaped mixing patterns are very similar: an upper cusp
  corresponds to a parabolic injection point on the boundary, while in
  the lower part of the pattern filaments are nicely packed in a
  parallel fashion. Although an annular unmixed region remains in the
  vicinity of the boundary, the partly-mixed pattern grows towards the
  boundary with time. This purely chaotic phase portrait is confirmed
  by the Poincar\'e section in (b) and (d), where a single trajectory
  fills the entire domain. In both cases two parabolic points can be
  inferred from the cusps in the upper and lower parts of the
  boundary. They correspond to separation points along whose unstable
  manifold fluid gets injected into the bulk, and to the corresponding
  reattachment at the opposite boundary.  The frame in (a) and (d)
  indicates where measurements are taken.  
}
\label{figs_closedhom}
\end{figure}

We first describe the essential features of the experimental set-up
\cite{Gouillart2007}. A cylindrical rod periodically driven on a
figure-eight path gently stirs viscous sugar syrup inside a closed vessel
of inner diameter $20~\text{cm}$
(Fig.~\ref{figs_closedhom}(a)). The fluid viscosity $\nu=5\times
10^{-4}\, \text{m}^2\cdot\text{s}^{-1}$ together with rod diameter
$\ell=16 \,\text{mm}$ and stirring velocity
$U=2~\text{cm}\cdot\text{s}^{-1}$ yield a Reynolds number $Re = U\ell/\nu
\simeq 0.6$, consistent with a Stokes flow regime. A spot of
low-diffusivity dye (Indian ink diluted in sugar syrup) is injected at
the surface of the fluid (Fig.~\ref{figs_closedhom:homsteps} (a)), and we
follow the evolution of the dye concentration field during the mixing
process (Fig.~\ref{figs_closedhom:homsteps}).  The concentration field is measured through the transparent
bottom of the vessel using a $12$-bit CCD camera at resolution $2000
\times 2000$. This protocol is a good candidate for efficient mixing: we
can observe on a Poincar\'e section (Fig.~\ref{figs_closedhom}(b)) ---
computed numerically for the corresponding Stokes flow \cite{Finn2001}
--- a large chaotic region spanning the \emph{entire} domain, including
the vicinity of the wall.

We also perform numerical simulations of ``dye homogenization'' for a
different stirring protocol. We consider a viscous version of the
blinking vortex flow~\cite{Aref1984,Jana1994}, that is, two
counter-rotating vortices alternatively switched on and off. Following
Jana et al.~\cite{Jana1994}, we study a realistic version of this
protocol consisting of two large fixed rods placed on a diameter of a
circular domain (see Fig.~\ref{figs_closedhom}(c)). To mimic the blinking
vortex, the two rods are rotated one after the other through angles
$\theta$ and $-\theta$, in a counter-rotating fashion. This stirring
protocol resembles the figure-eight, as the counter-rotating movement of
the vortices draws fluid from the boundary in some part of the domain
(the radial velocity $\vperp$ is positive), whereas it is pushed towards
the boundary in the other part ($\vperp<0$). The flow parameters are
$\theta = 270\degree$ (angular displacement of one rod at each half
period), $r=0.7$ (distance between the rods), $r_{\mathrm{inner}}=0.2$
(radius of the rods).  Length scale units are irrelevant here, and all
distances are scaled by the radius of the cylindrical vessel
$r_{\mathrm{outer}}=1$. In the same way, all times are rescaled by
the stirring period $\T$, so that we can work only with dimensionless
quantities in the following. A Poincar\'e section shows
(Fig.~\ref{figs_closedhom}(d)) that the chaotic region spans the entire
domain for this protocol as well. The evolution of a ``blob of dye'' is
mimicked by computing the positions of $10^6$ particles --- initialized
inside a small square in the center of the domain --- during 75 periods.

\begin{figure}
\begin{minipage}{0.49\columnwidth}
\subfigure[]{
\centerline{\includegraphics[width=0.99\textwidth]{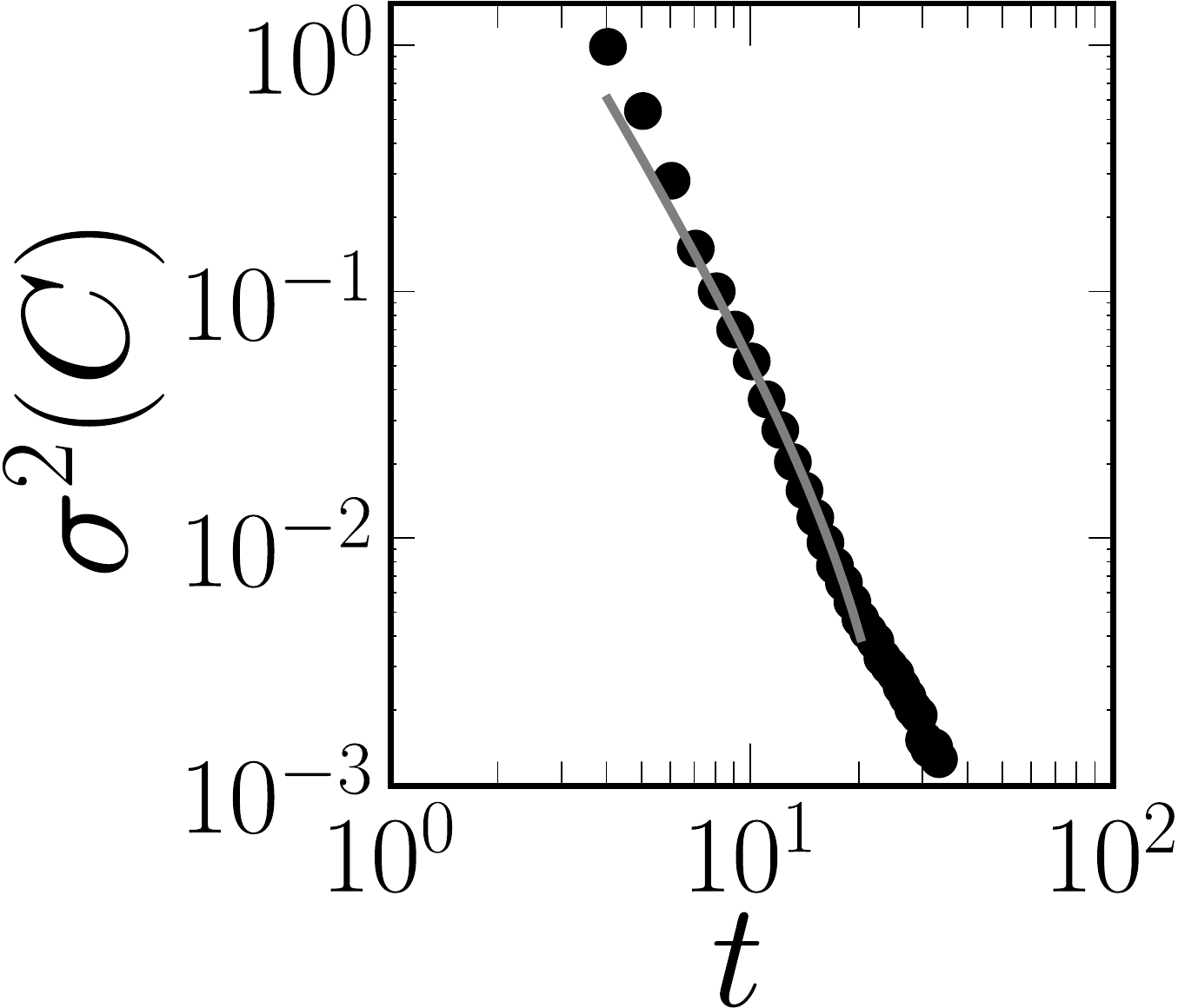}}
}
\end{minipage}
\begin{minipage}{0.49\columnwidth}
\subfigure[]{
\centerline{\includegraphics[width=0.99\textwidth]{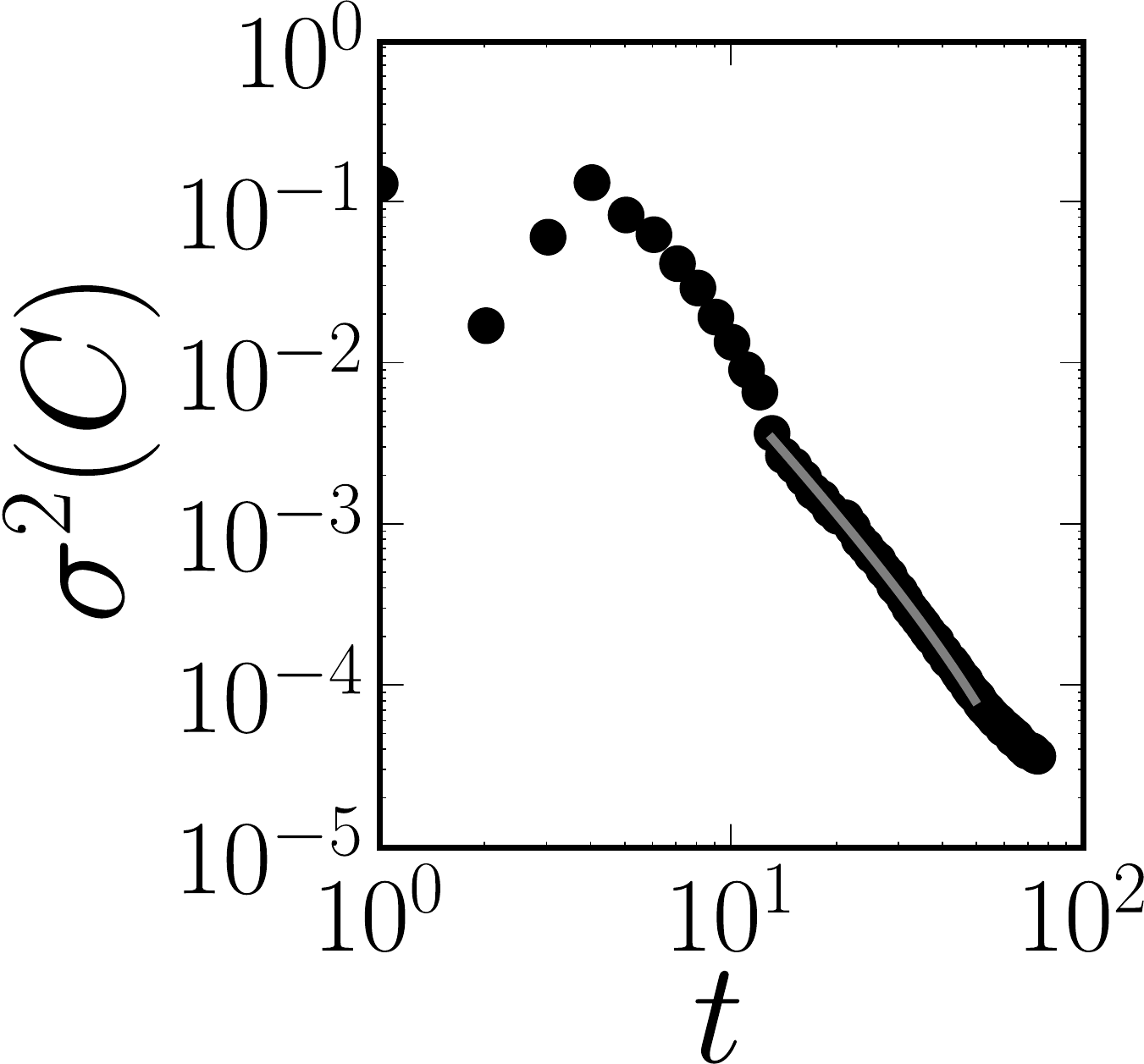}}
}
\end{minipage}
\subfigure[]{
\centerline{\includegraphics[width=0.99\columnwidth]{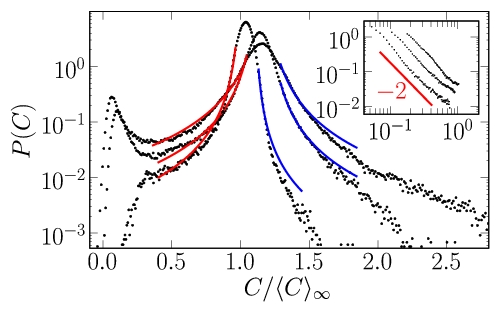}}
}
\caption{[Color online] Statistical properties of the concentration
  field measured in a central region (see frames in
  Fig.~\ref{figs_closedhom} (a) and (c)).  The concentration variance
  (black circles) is consistent with the evolution law of
  Eq. (\ref{eq:var_white}) (gray solid line) -- which is close to a
  power law -- both in (a) the figure-eight protocol and (b) the
  blinking vortex. The decay of variance is much slower than an
  exponential, which would be the signature of a strange eigenmode.
  (c) Concentration PDFs after 13, 17 and 31 stirring periods in the
  figure-eight protocol. Both sides of the peak can be fitted by power
  laws $(C_\mathrm{max}-C)^{-2}$ (see solid line fits, and
  inset). Inset: left (light-gray) tail of the peak, $P(C)$ against
  $|C_\mathrm{max}(t)-C|$. Also note the persistence of a white peak
  at $C=0$, which transforms into large shoulder for longer times.  }
\label{figs_closedhom:exp_results}
\end{figure}

In the experiments, we measure the concentration field inside a large
rectangle (see Fig.~\ref{figs_closedhom}(a)) far from the wall.  We
plot the resulting variance and probability distributions functions
(PDF) of the concentration in Fig.~\ref{figs_closedhom:exp_results}.
We observe a decay of the concentration variance $\sigma^2(C)$ that is
much closer to algebraic than exponential.  (see
Fig.~\ref{figs_closedhom:exp_results}(a)).  In fact, as we will see in
Sec.~\ref{sec:Cstat}, the decay is well approximated by
$\log\left(\tw/t\right)/t^2$. 
This behavior
persists until the end of the experiment (35 periods), by which time
the variance has decayed by more than three orders of magnitude. We
also note that PDFs of concentration have a wide shape, with power-law
tails on both sides of the maximum (see
Fig.~\ref{figs_closedhom:exp_results}(c)). Moreover, the PDFs of
concentration are highly asymmetrical. A persistent white peak at
zero-concentration values slowly transforms into a large shoulder at
weak concentration values. This implies that the light-gray wing of
the peak, corresponding to concentration values smaller than the most
probable value, is more important than the dark-gray wing on the other
side of the peak. Finally, the most probable value shifts slowly with
time towards lower values.

For the simulations, we compute a coarse-grained concentration field
in a large area far from the boundary (rectangle in
Fig.~\ref{figs_closedhom}(c)). The coarse-graining (0.01 here) scale
plays the role of the diffusive cut-off scale $\wb$. Again, we observe
an algebraic evolution of the concentration variance (shown in
Fig.~\ref{figs_closedhom:exp_results}(b)) that is well fitted by the same
decay law as in experiments.

\subsection{Hydrodynamics near the wall}
\label{sec:nearwall}

These experimental and numerical results are not
consistent with the exponential evolution of a single eigenmode of the
advection-diffusion operator.  In order to understand these scalings,
we first consider the various mechanisms at play. We will show that
the observed slow mixing arises from a subtle combination of
hydrodynamics, and the nature of the phase portrait at the wall.

As can be observed on the Poincar\'e section in
Fig.~\ref{figs_closedhom}(b), the chaotic region of the figure-eight
protocol spans the whole domain, and no transport barriers are
visible. (Elliptical islands can appear inside both loops of the
figure-eight for a smaller rod, but for a large enough rod we did not
detect such islands.) In particular, trajectories initialized close to
the wall boundary also belong to the chaotic region.  They eventually
escape from this peripheral region to visit the remainder of the phase
space, but only after a long time, as trajectories stick to the
no-slip wall.  This escape process takes place along the white cusp of
the heart-shaped mixing region, as can be seen in
Fig.~\ref{figs_closedhom}. This white cusp is bisected by the unstable
manifold of a separation 
point at the wall
(upper red dot in Fig.~\ref{figs_closedhom} (b) and (d)).  The
manifold divides trajectories reinjected from the left and right of
the separation point.  Since stretching is very weak close to the
wall, fluid drawn into the heart of the chaotic region from the wall
is poorly mixed at the moment when it is injected, as opposed to fluid
that has spent some time there already.  Because we inject the initial
blob of dye far from the boundary, poorly-stretched fluid injected
from the boundary into the heart of the chaotic region consists of
zero-concentration white strips that are interweaved into the mixing
pattern (see Fig.~\ref{figs_closedhom:homsteps} (b), (c) and (d)).

\mathnotation{\Af}{A}

We can better characterize such white strips in terms of hydrodynamics
near the no-slip wall.  Consider the velocity field~$\vb$ near the
vessel boundary. The wall can be treated as locally flat, and we define
local coordinates $\xpara$ and $\xperp$ that denote respectively
the distance along and perpendicular to the wall.  No-slip boundary
conditions impose $\vpara=0$ for $\xperp=0$ (on the wall) and the
corresponding first-order linear scaling for small $\xperp$,
\begin{equation}
  \vpara = \Af(\xpara)\,\xperp + \mathrm{O}(\xperp^2),
  \qquad \text{near the wall}.
  \label{eq_closed:DL}
\end{equation}
Note that we are modeling the \emph{net} velocity field, as evident
in the Poincar\'e sections in Figs.~\ref{figs_closedhom}(b)
and~\ref{figs_closedhom}(d), so we ignore the periodic
time-dependence.  Incompressibility implies
\begin{equation}
\frac{\partial \vpara}{\partial \xpara}
+ \frac{\partial \vperp}{\partial \xperp} = 0,
\label{eq_closed:incompr}
\end{equation}
which combined with~\eqref{eq_closed:DL} yields
\begin{equation}
  \vperp = -\tfrac{1}{2}\Af'(\xpara)\,\xperp^2  + \mathrm{O}(\xperp^3).
  \label{eq_closed:perp}
\end{equation}
Now from the Poincar\'e sections Figs.~\ref{figs_closedhom}(b) and~(d)
we can see that the only trajectories that consistently approach the
wall do so along a separatrix connected to the wall in the lower part
of the vessel (the lower dot in each figure).  All other trajectories
recirculate into the bulk.  The separatrix corresponds to a flow
re-attachment point on the boundary~\cite{Jana1994, Mezic2001,
  Haller2004}, which we refer to as \emph{parabolic points}.  (All
points on the boundary are parabolic fixed points, but the important
ones for us have separatrices emanating from them.  We only mean those
distinguished points when we refer to parabolic fixed points.)

If we choose~$\xpara=0$ to be the position of the lower separatrix,
then the velocity on the separatrix is
\begin{equation}
  \vpara \simeq 0,\qquad \vperp \simeq -\tfrac{1}{2}\A\,\xperp^2\,,
  \qquad \A \ldef \Af'(0),
  \label{eq_closed:sep}
\end{equation}
since~$\Af(0)=0$ in order that the separatrix and the wall be on the
same streamline.  Note that the linear part of the flow around the
re-attachment point vanishes, hence the name \emph{parabolic}. 
The requirement that particles approach the wall
along the separatrix implies~$\A>0$.
%
%
Integrating~$\dot{x}_\perp = \vperp$ using Eq.~\eqref{eq_closed:sep},
we find
\begin{equation}
  \xperp(t)=\frac{x_0}{1 + \A tx_0},
\label{eq_closed:xperp}
\end{equation}
where~$x_0$ is the initial~$\xperp$ coordinate of the particle.
Equation~\eqref{eq_closed:xperp} predicts that the distance~$d(t)$
between the wall and a particle on the lower separatrix shrinks as
\begin{equation}
  d(t)\simeq {1}/{\A t},\qquad t \gg 1/\A x_0.
  \label{eq:d_manip}
\end{equation} 
This scaling was already derived in \cite{Chertkov2003a}, from the
same dimensional reasoning.  The rate of approach along the separatrix
constrains the approach to the wall of the entire mixing pattern.  We
verified both in the experiment and in the simulation that $d(t)$ is
indeed well-approximated by a power-law scaling $d(t)\sim 1/t$.  Note
also that Eq.~\eqref{eq:d_manip} implies that particles along the
separatrix `forget' their initial condition for long times.  This can
be seen in Figs.~\ref{figs_closedhom}(a) and~(c): material lines
`bunch up' against each other in the lower part of the domain faster
than they approach the wall.

To ensure mass conservation, a quantity of unmixed white fluid scaling
as $\dotd(t) \propto t^{-2}$ is injected periodically in the mixing
pattern. As each newly injected white strip has approximately the same
length (determined by the extent of the rod path), the
width~$\stripw(t)$ of a strip injected at time $t$ must also scale as
\begin{equation}
  \stripw(t) \ldef \lvert\dotd(t)\rvert = 1/\A t^2,
  \label{eq:stripwdef}
\end{equation}
where time $t$ has been rescaled by the period $\T=1$.  

The origin of our slow scaling now emerges.  Clearly, the mixing
pattern is chaotically stretched and folded by the rod at each
half-cycle, in the same manner as in a baker's map.  Yet the folds are
not stacked directly onto each other but are \emph{interweaved with
  the most recently injected white strip}.  Since each new white strip
has a large width that decreases only algebraically with time, the
decay of concentration is slowed down by this injection of unmixed
material.

The dominant mechanism for mixing can be summed up as follows: (i)
chaotic stretching imposes that the typical width of a filament of dye
in the bulk (\ie~far from the wall) shrinks exponentially down to the
diffusion or measurement scale; yet (ii) wide strips of unmixed fluid
of width $\stripw(t) \propto t^{-2}$ are periodically interweaved with
these fine structures.  Both protocols have in common a chaotic region
that spans the entire domain, which imposes the presence of parabolic
separation points on the boundary~\cite{Jana1994, Mezic2001,
  Haller2004}.  In the next section, we generalize the baker's map
model to include such a parabolic point at the boundary, and reproduce
the dominant features observed experimentally and numerically.

\subsection{A modified baker's map model}
\label{sec:modbaker}

We now wish to derive quantitative predictions to explain the observed
algebraic scaling for the concentration variance.  In the same spirit
as in Section~\ref{sec:SEhyp}, we simplify the two-dimensional problem
by characterizing only one-dimensional concentration profiles $C(x,t)$
perpendicular to the stretching direction along which dye filaments
align.  The effect of the mixer during a half-period boils down to the
action of a one-dimensional map that transforms concentration profiles
by interweaving an unmixed strip of fluid with two compressed images
of the profile.  The width of each decays in time as $\stripw(t) \propto
t^{-2}$, owing to the parabolic point on the boundary
(Section~\ref{sec:algebraic}).  We therefore mimic the behavior of
our mixer with a one-dimensional map $\g$, in the spirit of the
baker's map.

The map $\g$ is defined on $[0,1]$ for simplicity. It evolves concentration
profiles as in Eq.~\eqref{eq:Cevolve} and satisfies the following: (i) it is a
continuous one-to-two function, to account for the stretching and folding
process; (ii) the `wall' at $x=0$ is a
marginally unstable (i.e.  parabolic) point of $\g^{-1}$, so that the correct
dynamics are reproduced by expanding $\g^{-1}(x)\simeq x + ax^2 + \cdots$
($\A>0$), for small $x$; (iii) because of mass conservation, at each $x$ the
local slopes of the two branches, $g_1$ and $g_2$, of~$\g$ add up to $1$.

Other details of $\g$ are unessential for our discussion.  As in
Section~\ref{sec:algebraic}, diffusion is mimicked by letting the
concentration profile diffuse between successive iterations of the map
(with no-flux boundary conditions).  This model is a modified baker's
map, with a parabolic point at $x=0$, as opposed to the 
baker's map in Section~\ref{sec:algebraic}, where the dynamics
are purely hyperbolic. The expression of $\g_1$ close to $x=0$ assures
that the distance between the origin and successive iterates of a point
by $\g_1$ shrinks as $d(t)\simeq 1/at$, which is the same as
Eq.~(\ref{eq:d_manip}) obtained in the experiments with no-slip
hydrodynamics.

\begin{figure}[t]
\begin{center}
\subfigure[]{
\includegraphics[width=8cm]{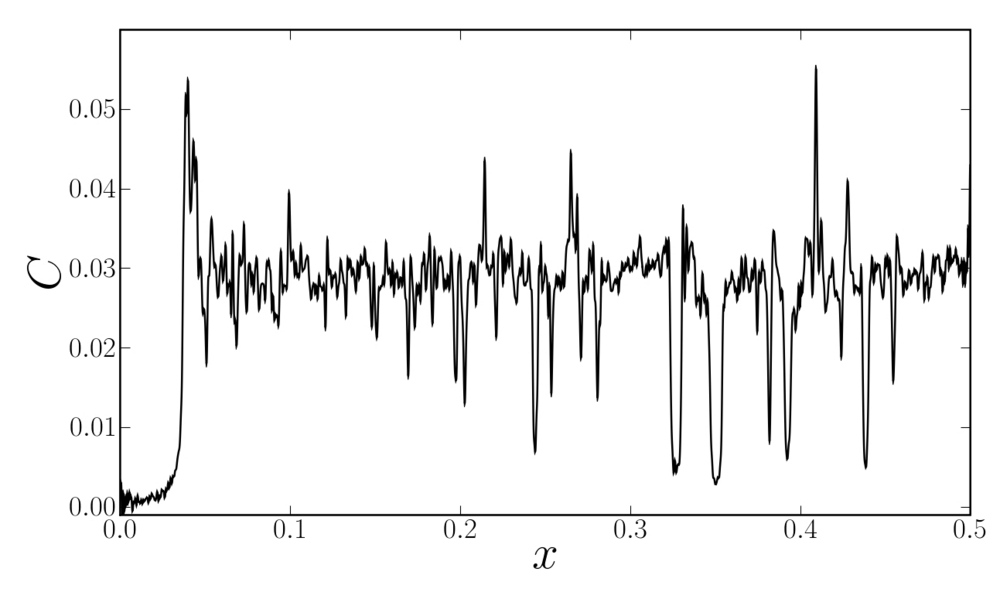}
}
\subfigure[]{
\includegraphics[width=8cm]{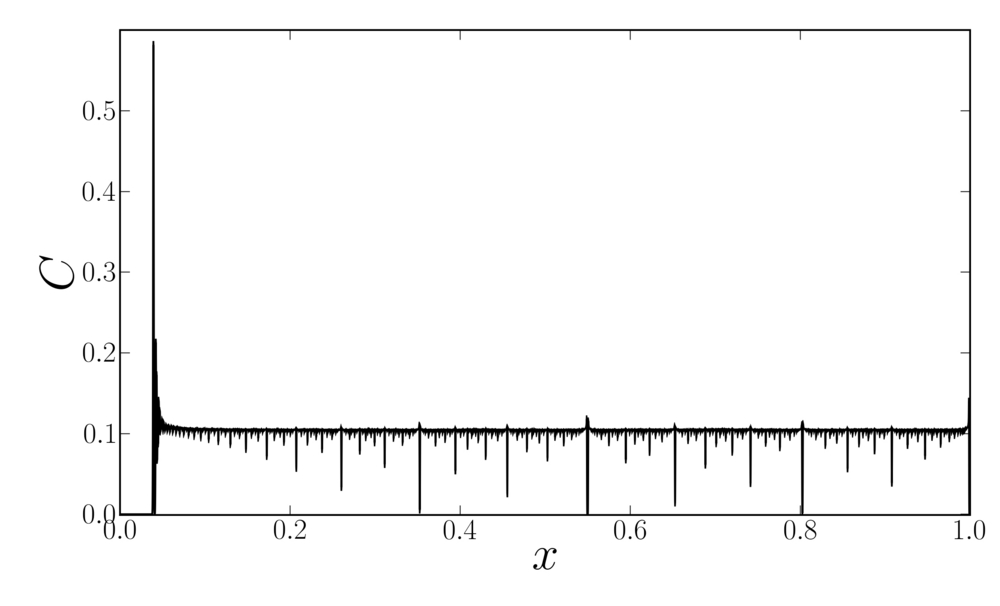}
}
\subfigure[]{
\includegraphics[width=8cm]{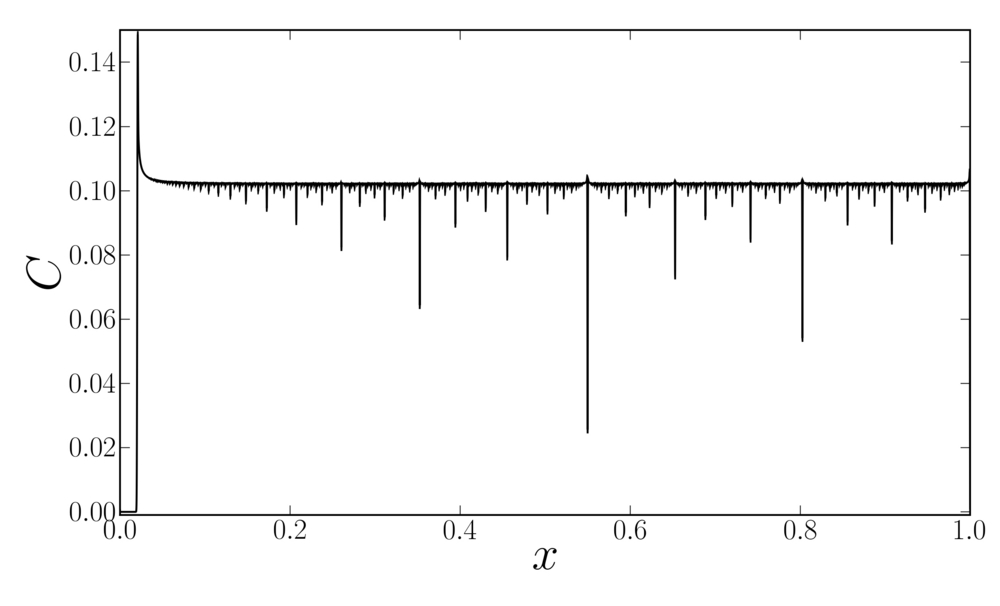}
}
\end{center}
\caption{Concentration profiles from (a) the figure-eight experiment
  after 13 stirring periods, and from the 1-D model after (b) 25 and
  (c) 50 iterations of the map.  Note the presence of ``white fluid''
  ($C=0$) near $x=0$ in all cases.}
\label{figs_closedhom:profs}
\end{figure}

We numerically evolve concentration profiles for the specific choice
for $\g$,
\begin{equation}
\left\{\begin{array}{cc}
\g_1(x)&= x - \A x^2 + (\gamma -1 + \A) x^3, \\
\g_2(x)&= 1 - \A x^2 + (\gamma -1 + \A) x^3,
\end{array}
\right.
\label{eq:g_def}
\end{equation}
with $\gamma=0.55$ and $\A=0.9$. We fix $\kappa=10^{-7}$.  In our map
$\g_1(1)=\g_2(1)=\gamma$, and as for the baker's map we approximate by
$\gamma$ the mean stretching realized by $\g_1$, and by $(1-\gamma)$
the mean stretching realized by $\g_2$ --- although stretching is not
constant along the two branches. We choose $\gamma \neq 0.5$ for an
uneven stretching in the bulk, as in the experiments where fluid
particles that stay close to the rod for long times experience more
stretching than particles left behind. Our initial condition is of the
form~\eqref{eq:C0}, with~$x_a = (1-\delta)/2$ and~$x_b =
(1+\delta)/2$, i.e., a strip of width~$\delta$ centered on~$x_c=1/2$.

Figure~\ref{figs_closedhom:profs} shows concentration profiles obtained
after several iterations of the map. Strong similarities are observable
between concentration profiles obtained in the experiment
(Fig.~\ref{figs_closedhom:profs}(a)) and in the map
(Fig.~\ref{figs_closedhom:profs}(b)--(c)). In both cases a thin layer of
``white fluid'' ($C=0$) is present near $x=0$. Its width decreases as $1/\A
t$ due to the parabolic point on the boundary. In the experiment, the
concentration pattern in the bulk (far from the wall) is characterized by
sharp spikes at zero or low concentration values, whereas fluctuations
are quite weak elsewhere. The sharp spikes correspond to white strips
recently injected from the boundary into the bulk. For the map, the bulk
pattern (far from~$x=0$) is clearly dominated by a set of thin spikes,
which are recently injected white strips. These white strips are images
of the boundary region at $x=0$ by $\g_2$, which are successively iterated by
$\g_1$ or $\g_2$ after their injection in the bulk.

\begin{figure}
\subfigure[]{
\centerline{\includegraphics[width=0.95\columnwidth]{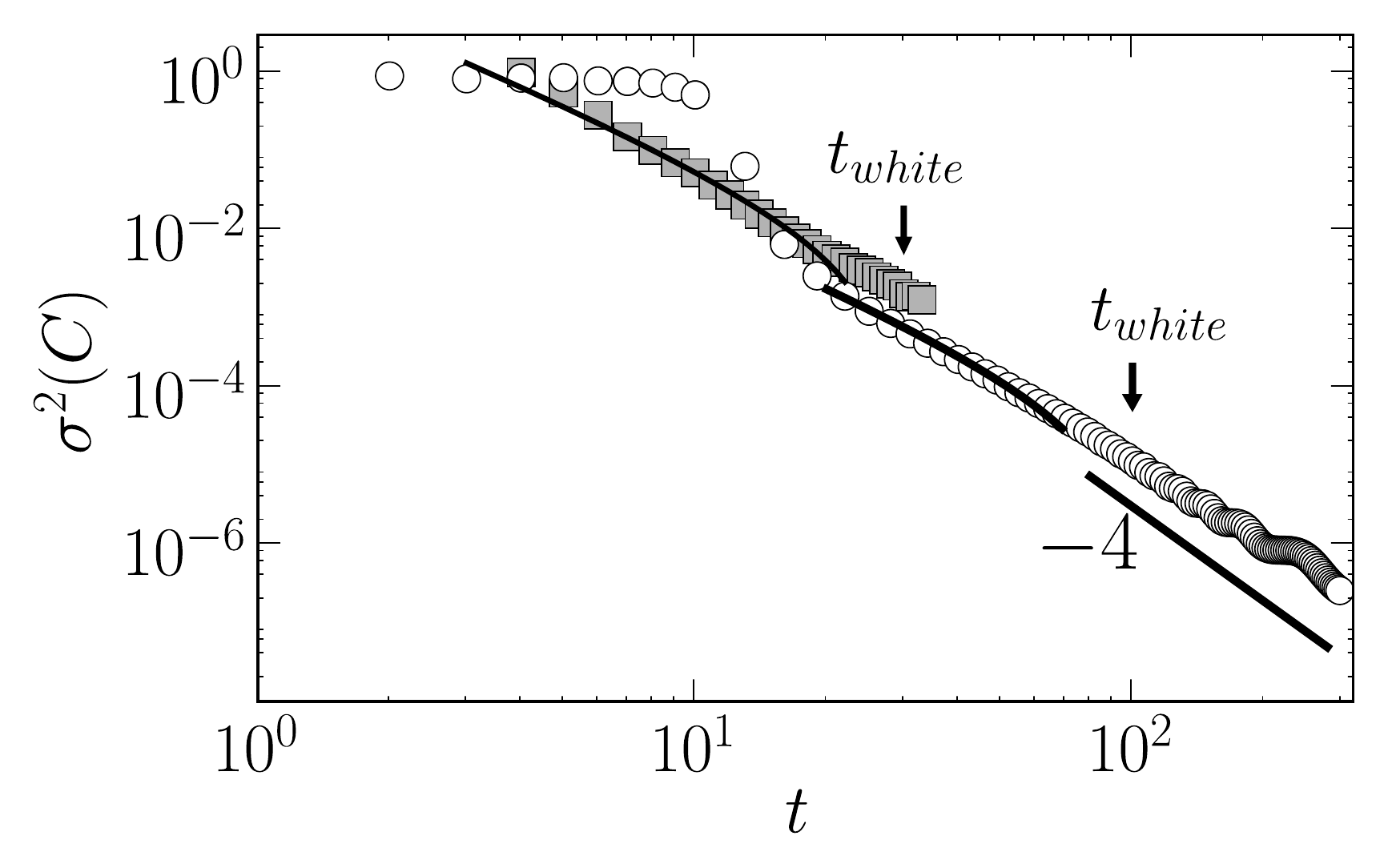}}
}
\subfigure[]{
\centerline{\includegraphics[width=0.95\columnwidth]{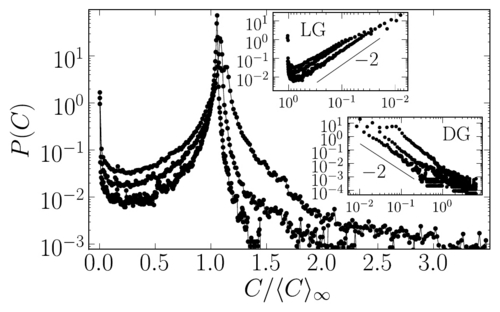}}
}
\caption{Variance and PDFs of the concentration field measured in a
  central region for a blob of dye transformed by the modified baker's
  map (\ref{eq:g_def}). (a) The concentration
  variance (circles) shows an evolution close to a power-law,
  comparable to the figure-eight experiment (square symbols).  The
  solid line fit and the $t^{-4}$ slope correspond respectively to the
  evolution dictated by Eqs. (\ref{eq:var_white}) and
  (\ref{eq:var_gray_aftertw}). (b) The concentration PDFs have wide
  power-law tails on both sides of the peak.  The dark-gray tail
  corresponding to high concentration values decays much faster than
  the light-gray one (weak concentration values), since the latter is
  continuously fed by the injection of white fluid from the boundary.}
\label{figs_closedhom:model_results}
\end{figure}

The suitability of our model is also strengthened by statistical
properties of the concentration field, which closely resemble the
experiment. Figure~\ref{figs_closedhom:model_results}(a) shows the
concentration variance for the map (measured in a central region)
superimposed with experimental data: again we find an algebraic
evolution. Moreover, there is a strong similarity between the
concentration PDFs depicted in
Fig.~\ref{figs_closedhom:model_results}(b) and the experimental ones
shown in Fig.~\ref{figs_closedhom:exp_results}(c).  In particular,
they both exhibit power-law tails.  We will see in
Section~\ref{sec:Cstat} that our modified baker's map is simple enough
for the concentration statistics to be calculated explicitly.

\section{Concentration statistics for the modified baker's map}
\label{sec:Cstat}

The simplicity of the model introduced in Section~\ref{sec:modbaker}
allows us to calculate the statistical properties of the concentration
field. In this section, our interpretation is based mostly on our
simplified model, although comparisons with the experiment are also
made. We first consider the simple case of an initially-uniform blob,
for which we characterize the concentration pattern by counting
iterates of injected white strips, since these dominate the
concentration pattern. We will treat more general initial conditions
in Section~\ref{sec:discuss}.  We focus here on a central region where
the concentration profile is at least partly mixed --- that is, away
from $x=0$, where~$x$ is the map coordinate.  The concentration PDFs
and variance presented above have for example been measured in the
range $x \in [0.2,0.9]$ in the map. We have checked that the variance
measured in the whole domain evolves trivially as $1/t$, as it is
dominated by the remaining white pool at the wall. (The variance in the
numerical simulations of Salman and Haynes is measured in the whole
domain and displays a $t^{-1}$ evolution during the algebraic phase
\cite{Salman2007}.)  Here we are interested in the more complex
evolution in the bulk, where stretching is high and the pattern seems
``well mixed'' after a few periods.

The modified baker's map of Section~\ref{sec:modbaker} transforms an
initial blob of dye of width $s_0$ into an increasing number of strips
with widths $s_0 \lac_1 \cdots \lac_t$, resulting from different
stretching histories inside the mixed region, where $\lac_t$ is the
compression experienced at time $t$. White strips also experience this
multiplicative compression
starting from their injection time. Because of diffusion, a strip of
dye or white fluid is only compressed down to the local diffusive
Batchelor scale $\wb$, that we approximate by~$\wb =
\sqrt{{\kappa}/{(1-\Gamma^2)}}$, where
\begin{equation}
\log\Gamma=-\left\langle
\log
\left|\frac{\partial g^{-1}(x)}{\partial x}\right|\right\rangle
\end{equation}
is the Lyapunov exponent and $\langle\cdots\rangle$ is the spatial mean
taken over the region of measurement.  
For a moderate
stretching inhomogeneity in the bulk, we expect $\Gamma$ to be close
to $1/2$.  In experiments as in simulations, we probe the
concentration field on a pixel size, or box size, which is smaller
than $\wb$.

As in the baker's map, different values of $C$ correspond to a
different combination of superimposed strips in a box of size $\wb$.
We characterize $P(C)$ by considering the different widths of
reinjected white strips that one can find in a such a box. We will
distinguish between three generic cases corresponding to a partition
of the histogram $P(C)$ in three different regions (see
Fig.~\ref{figs_closedhom:exp_results}(c) and
Fig.~\ref{figs_closedhom:model_results}(b)): a white (W) peak at $C=0$
corresponding to recently-injected white strips that are still wider
than $\wb$, and light-gray (LG) and dark-gray (DG) tails corresponding
to respectively smaller and larger concentrations than the peak (mean)
concentration. Once we have quantified the proportion of boxes
contributing to these different values of $C$, the variance is readily
obtained as
\begin{equation}
  \sigma^2(C)=\int (C-\langle C \rangle)^2 P(C)\, \d C=
  \sigma^2_{\mathrm{W}} + \sigma^2_{\mathrm{LG}} + \sigma^2_{\mathrm{DG}}.
  \label{eq_closedhom:var}
\end{equation}
We treat
each region of the histogram in turn in Sections~\ref{sec:W}--\ref{sec:DG},
and combine the results in Section~\ref{sec:alltogether}.

\subsection{White pixels}
\label{sec:W}

Let us start with white (zero) concentration values that come from the
stretched images of white strips injected before time~$t$. White
strips injected at an early time have been stretched and wiped out by
diffusion, that is their width has become smaller than $\wb$. A white
strip injected at time $\tnot$ has been compressed to a width
$\stripw(\tnot)\lac_{\tnot+1}\cdots\lac_t$ at time $t$.
We neglect
the spatial variation of $\lac(x)$ in the bulk and approximate
$\lac_{\tnot+1}\cdots\lac_t \simeq \Gamma^{t-\tnot}$. The oldest white
strips that can be observed have been injected at time $t=\ti(t)$,
where~$\ti(t)$ is defined by
\begin{equation}
  \stripw(\ti) \, \Gamma^{t-\ti}=\wb\,.
  \label{eq:tidef}
\end{equation}
Note that $t -\ti(t)$ -- that is the number of periods needed to compress an injected strip
to $\wb$ -- is a decreasing function of time. 
After a time~$t=\tw$ defined by
\begin{equation}
  \stripw(\tw) = \wb\,,
\end{equation}
the injected white strip is smaller than $\wb$ and no white pixels can
be observed. For $t < \tw$, we can observe all white strips that are
images of strips injected between $\ti(t)$ and $t$, and the number of
white pixels is proportional to
\begin{equation}
  n_{\mathrm{W}} = \sum_{n=\ti(t)+1}^{t} \stripw(n)
  =  d(t) - d(\ti(t)),
\end{equation}
where~$\stripw(t) = d(t)-d(t+1)$. 

\begin{figure}
\subfigure[]{
\includegraphics[width=8cm]{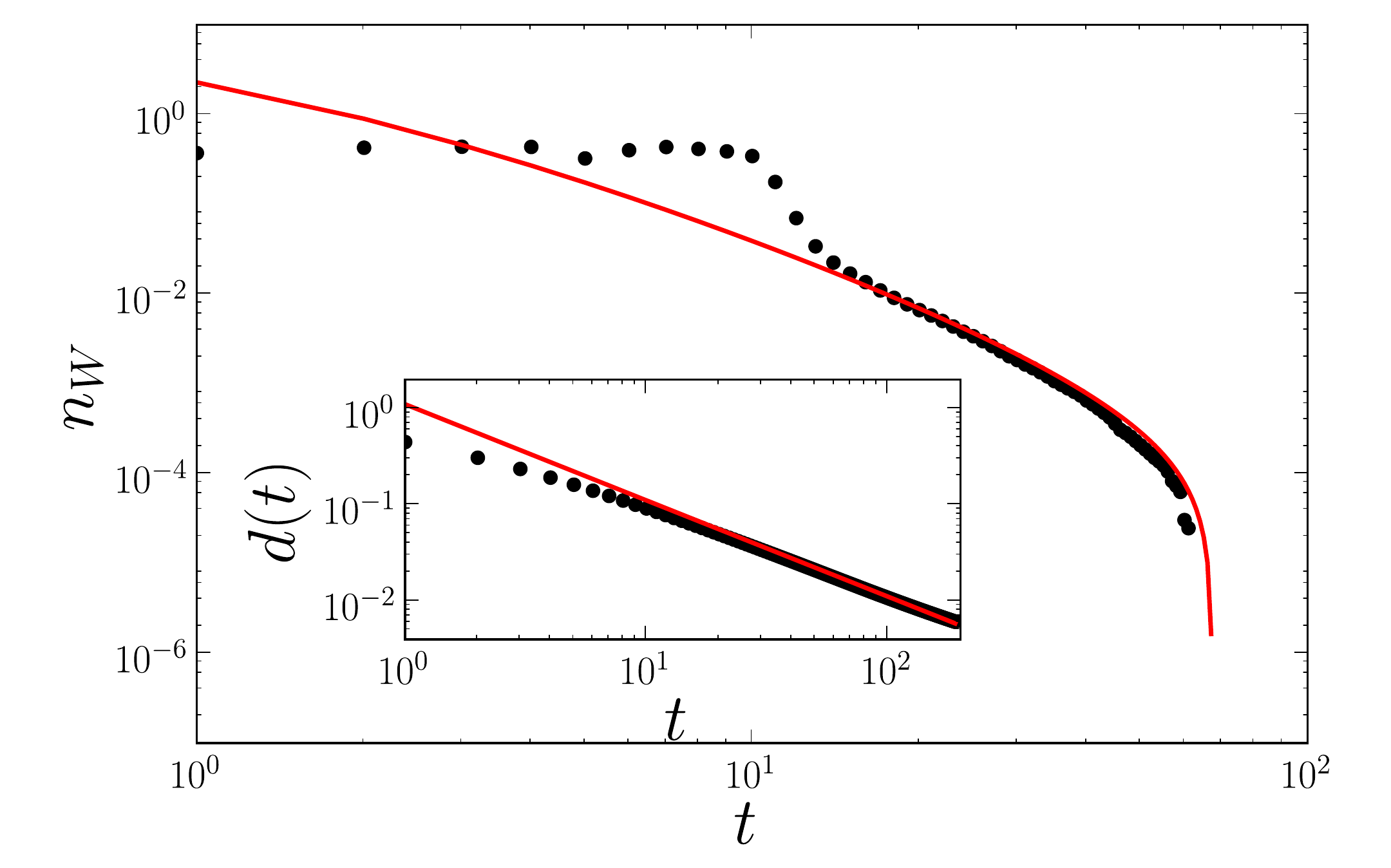}
}
\subfigure[]{
\includegraphics[width=8cm]{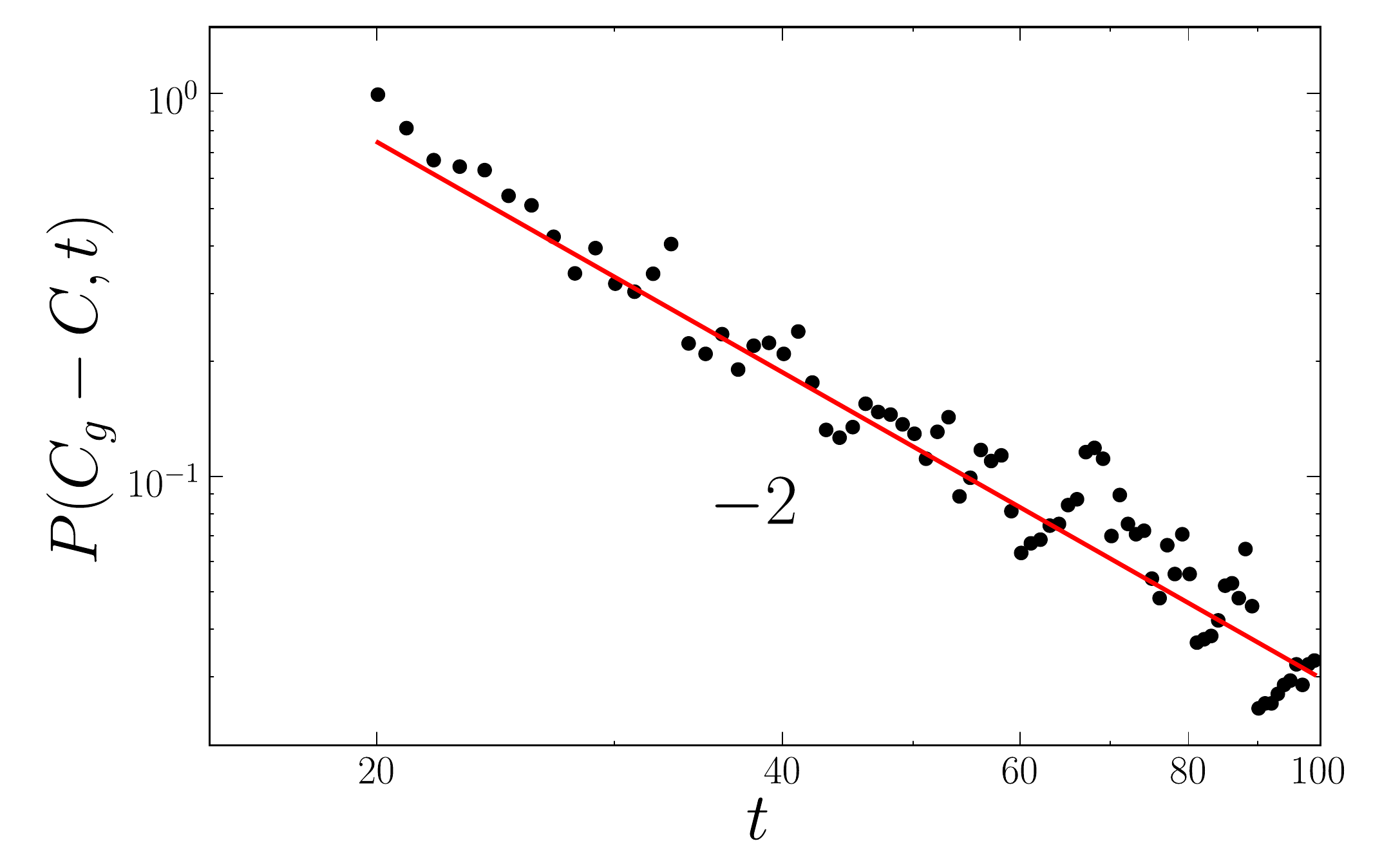}
}
\caption{Simulations of dye homogenization by the modified baker's
  map: (a) Fraction of white pixels (where $C=0$) $n_{\mathrm{W}}$ as a function
  of iterate (full circles). At early times diffusion has not yet
  started to broaden strips of dye, and $n_{\mathrm{W}}$ remains approximately
  constant. When $n_{\mathrm{W}}$ begins to decay, it closely follows
  Eq.~\eqref{eq_closedhom:nw} (solid line) obtained by counting the
  images of injected white strips that have not yet been compressed
  down to the diffusion scale $\wb$. Inset: the distance between the
  dye pattern and the wall (measured by the position of the first peak
  in Fig.~\ref{figs_closedhom:profs}) evolves as $d(t)=1/at$ (solid
  line). (b) Probability of a light-gray concentration level with a
  given distance to the peak, i.e. probability that $\left| C_\mathrm{g}
- \mathrm{d}C -C \right| \leq \epsilon $
($C_g\sim 0.1$, $\mathrm{d}C=3\times
  10^{-2}$, $\epsilon=10^{-3}$, 
as a function of time.
  $P(C_\mathrm{g}-C)$ agrees well with the $t^{-2}$ evolution predicted by
  Eq.~\eqref{eq_closedhom:gt} with~$\Gamma \simeq 1/2$ (solid curve).}
\label{figs_closedhom:white}
\end{figure}

We now use the expression for $d(t)$ imposed by dynamics close to the
parabolic point, $d(t) \simeq 1/at$, which yields for large $t$
\begin{equation}
  n_{\mathrm{W}} \simeq (t -\ti)(a \ti t)^{-1}.
  \label{eq:nW}
\end{equation}
From the definition~\eqref{eq:tidef} of the injection time $\ti(t)$,
\begin{equation}
  t-\ti \simeq (\log{\A t^2}+\log{\wb})/\log{\Gamma}
\end{equation}
for large $t$, therefore
\begin{equation}
  n_{\mathrm{W}} \simeq (\log{\A t^2}+\log{\wb})/(\A t^2\log{\Gamma}).
  \label{eq_closedhom:nw}
\end{equation}
The
fraction of white pixels $n_{\mathrm{W}}$ is plotted versus time in
Fig.~\ref{figs_closedhom:white}(a). (There are no free parameters.)
For later times, we find an excellent agreement between the data and
the expression~\eqref{eq_closedhom:nw} for $n_{\mathrm{W}}$. Note that
during the first few iterations $n_{\mathrm{W}}$ is constant: this
corresponds to the initial phase when dye strips are still wider than
$\wb$ and diffusion is ineffective (i.e. up to $t$ such that
$\delta\times \Gamma^t=\wb$). The concentration variance is also
almost constant during this initial phase, so we discard it. We deduce
the contribution of the white pixels to the concentration variance for
$t<\tw$,
\begin{equation}
\sigma^2_{\mathrm{W}}=n_{\mathrm{W}} \langle C \rangle^2 \simeq
\delta^2\times\frac{\log{\A t^2}+\log{\wb}}{\A t^2\log{\Gamma}}.
\label{eq:var_white}
\end{equation}
Of course, $\sigma^2_{\mathrm{W}}=0$ for $t>\tw$, since by then there are no
purely-white strips left.

\subsection{Light-gray tail}
\label{sec:LG}


We now focus on the distribution of light-gray values corresponding to white
strips that have just been compressed below the cut-off scale $\wb$. A white
strip is first injected between images of the mixing pattern where
fluctuations are lower (see Fig.~\ref{figs_closedhom:profs}). Fluctuations
measured in a pixel are therefore mostly due to a recently-injected white
strip that is superimposed onto a homogeneous distribution.  We approximate
the measured value $C$ as the average of the largest white strip with width
$\lw < \wb$, and mixed ``gray'' fluid whose concentration is close to the
most probable concentration $C_{\mathrm{g}}$. A box
containing a white strip of scale $\lw$ 
thus carries a concentration
\begin{equation}
  C = \Cg \times (1-\lw/\wb),
  \label{eq:Clw}
\end{equation}
and we can relate the concentration PDF $P(C)$ to the
distribution~$Q(\lw)$ of widths of images of the injected white strips in
the following way:
\begin{equation}
  \PC = Q\big(\lw\big)\l\lvert
  \frac{d\lw}{dC}\r\rvert
  = \frac{\wb}{\Cg}\, Q\big(\lw\big). 
  \label{eq:Cvslambda}
\end{equation}
$Q(\lw)$ is easily retrieved from standard combinatorial arguments. A
white strip injected at $\tnot$ is transformed into $2^{t-\tnot}$ images
with scales $\stripw(\tnot) \Gamma^{t-\tnot}$ (once again we consider only
the mean stretching $\Gamma$, which amounts to matching a given
concentration to a unique injection time). In a ``quasi-static''
approximation, we neglect the algebraic dependence of $\lw$ (and hence
of $C$ as well) on $\tnot$ in the factor $\stripw(\tnot)$ compared to the exponential
dependence in $\Gamma^{t-\tnot}$. Therefore
\begin{equation}
Q(\lw)=(\lw/\stripw(t_{0}))^{\log(2)/\log(\Gamma)}
\times (1/\lw \log{\Gamma}),
\end{equation}
resulting in

\begin{align}
\PC &=
[\stripw(t)]^{\,\log{2}/\log{(\Gamma^{-1})}}\frac{\wb}{C_{\mathrm{g}}(t)}
\bigg[\wb \l(1 - \frac{C}{\Cg}\r) \bigg
]^{\frac{\log{2}}{\log{\Gamma}}-1}\nonumber \\
&= \PT \bigl[\Cg - C \bigr]^{(\log{2}/\log{\Gamma})-1}\;.
\label{eq:PC}
\end{align}
$\PC$ thus has a power-law tail in the light-gray levels whose exponent
depends on the mean stretching $\Gamma$. We observe satisfactory agreement
between this prediction and both experimental data and numerical 1-D
simulations (see Figs.~\ref{figs_closedhom:exp_results}(c)
and~\ref{figs_closedhom:model_results}(b)). Indeed, for the tail in
Fig.~\ref{figs_closedhom:exp_results}(c) and 
Fig.~\ref{figs_closedhom:model_results}(b) we measure $\PC\propto
(C-\Cg)^{-\alpha}$ with $\alpha \lesssim 2$, consistent with
$\Gamma \lesssim 1/2$, a fairly homogeneous stretching, as expected
for~$\gamma=0.55$. Also note that
the amplitude of the light-gray tail decreases with time as a power law,
\begin{equation}
\PT\propto [\stripw(t)]^{\,\log 2/\log(1/\Gamma)} \propto
t^{-2(\log
2/\log(1/\Gamma))}.
\label{eq_closedhom:gt} 
\end{equation}

We have plotted in Fig.~\ref{figs_closedhom:white}(b) the probability of a
concentration value at fixed distance from the maximum. The observed
evolution scales as a power-law $t^{-2}$, as expected from our calculation. We
deduce the contribution of light-gray pixels to the concentration
variance,
\begin{equation}
\sigma^2_{\mathrm{LG}}=\PT\int_{C_{\mathrm{min}}}^{C_{\mathrm{g}}}
(C_{\mathrm{g}}-C)^{2-\alpha(\Gamma)} \d C\,,
\end{equation}
where $\alpha(\Gamma)=1-\log{2}/\log{\Gamma}$, and $C_{\mathrm{min}}$ is the
smallest concentration observed ($C_{\mathrm{min}}=0$ for $t<\tw$ and
$C_{\mathrm{min}}=\Cg(1-\stripw(t)/\wb)$ for $t>\tw$). For $t<\tw$
the integral is constant and $\sigma^2_{\mathrm{LG}}\propto \PT \propto \stripw(t)
\propto t^{-2}$. On the other hand, for $t\ge\tw$,
\begin{equation}
  \sigma^2_{\mathrm{LG}}=\frac{\PT}{2+\alpha(\Gamma)}
	[\Cg-C_{\mathrm{min}}]^{\,3-\alpha(\Gamma)}
	\propto t^{-\l(6+2\frac{\log 2}{\log \Gamma}\r)}\,.
\label{eq:var_gray_aftertw}
\end{equation}
For $\alpha(\Gamma)\sim 2$ as we observed, the exponent in the above power law
is about $-4$.

\subsection{Dark-gray tail}
\label{sec:DG}


Let us now turn to the dark-gray part of the PDF. In our case, high
concentration values correspond to black strips of dye that have experienced
little compression, so that they have not been grayed-out by averaging
with many other strips. This time, it is not sufficient to consider only the
mean stretching $\Gamma$ to characterize such strips as we did in
Sections~\ref{sec:W} and~\ref{sec:LG}, since stretching histories far from the
mean are involved. Looking at the concentration profiles in
Fig.~\ref{figs_closedhom:profs}, we observe that the highest concentration
values come from the reinjection of black strips pushed to the pattern
boundary where they have experienced lower stretching than inside the pattern
core. Such a positive concentration fluctuation is then mixed with the
remainder of the pattern as successive images are compressed by a factor of
order $\Gamma$, in the same way as injected white strips. Many images of the
initial blob may have aggregated inside a box of size $\wb$. If the decay of
this highest-concentration ``cliff'' is slower than $\Gamma^t$ --- the decay
of an injected fluctuation inside the bulk --- we can apply the same method
for computing the shape of the dark-gray tail as we did for white strips and
the light-gray tail.

In the spirit of Eq.~\eqref{eq:Clw}, we write 
\begin{equation}
  C = \Cg \times (1-\lb/\wb) +\lb/\wb,
\end{equation}
and relate the width $\lb$ to the injection time $\tnot$ as in
Section~\ref{sec:LG}.  This leads again to a power-law
dependence~$(C-\Cg)^{-2}$, this time for the dark-gray tail. This is in good
agreement with the observed scalings for both experimental and numerical PDFs
(see Figs.~\ref{figs_closedhom:exp_results}(c)
and~\ref{figs_closedhom:model_results}(b)).

We now wish to estimate the time decay of the amplitude of this tail.
To do so, we evaluate the amplitude of the highest concentration
fluctuation, located on the left boundary of the pattern (see
Fig. \ref{figs_closedhom:profs}(b) and (c)). This will give us the
concentration value for which a number of boxes of order 1 contribute
to the histogram, and hence provide an approximation of the amplitude
of the tail.  The contribution of the dark-gray tail is tiny compared
to the light-gray one, since, after a few periods, only a few boxes of
size $\wb$ on the border of the pattern have an amplitude
significantly greater than the mean (see
Figs.~\ref{figs_closedhom:profs}(b) and (c)), whereas the width of the
remaining white pool is much larger. Moreover, we show below that this
amplitude decays faster than the contribution of white strips.

\begin{figure}
\subfigure[]{
\centerline{\includegraphics[width=8cm]{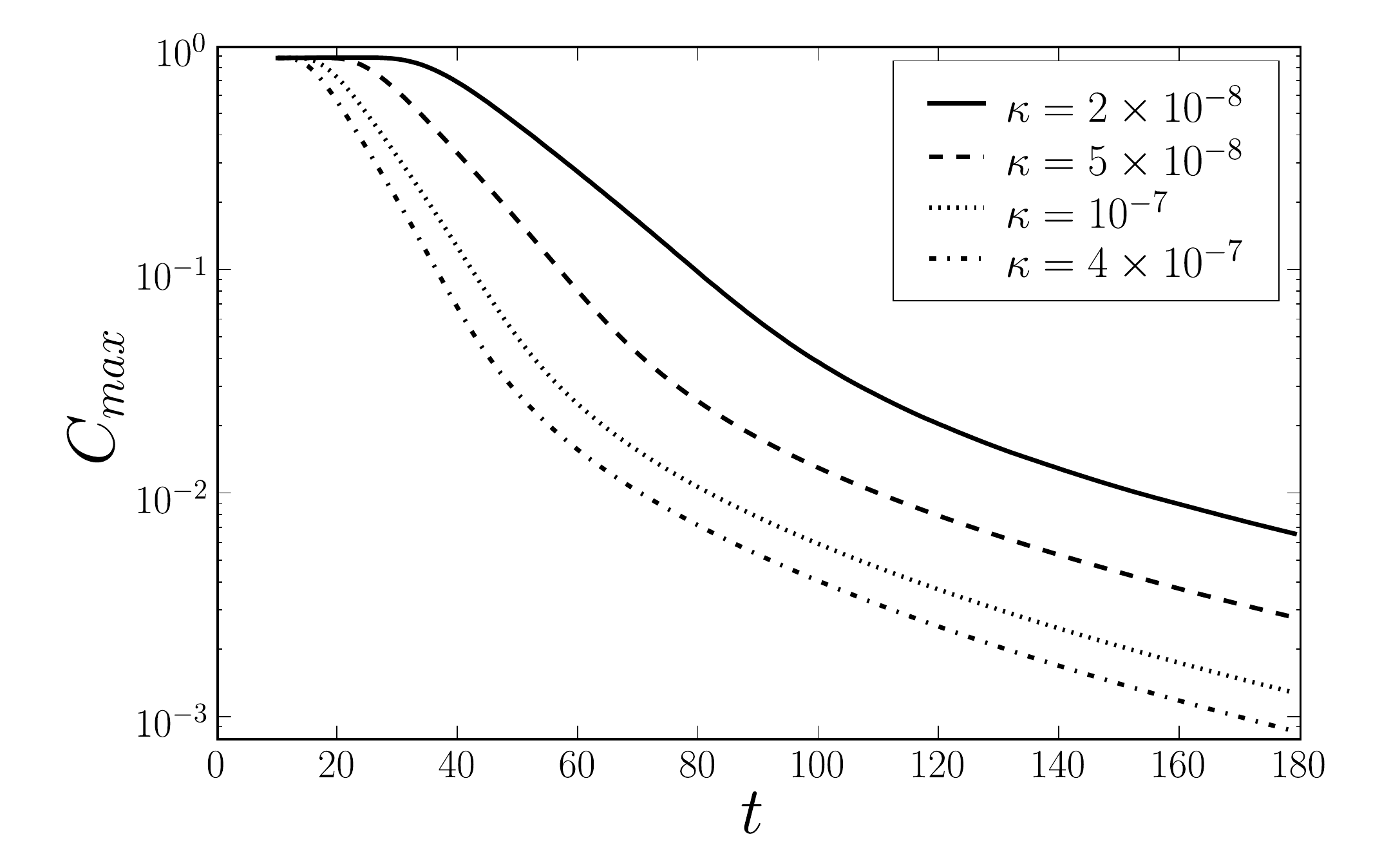}}
}
\subfigure[]{
\centerline{\includegraphics[width=8cm]{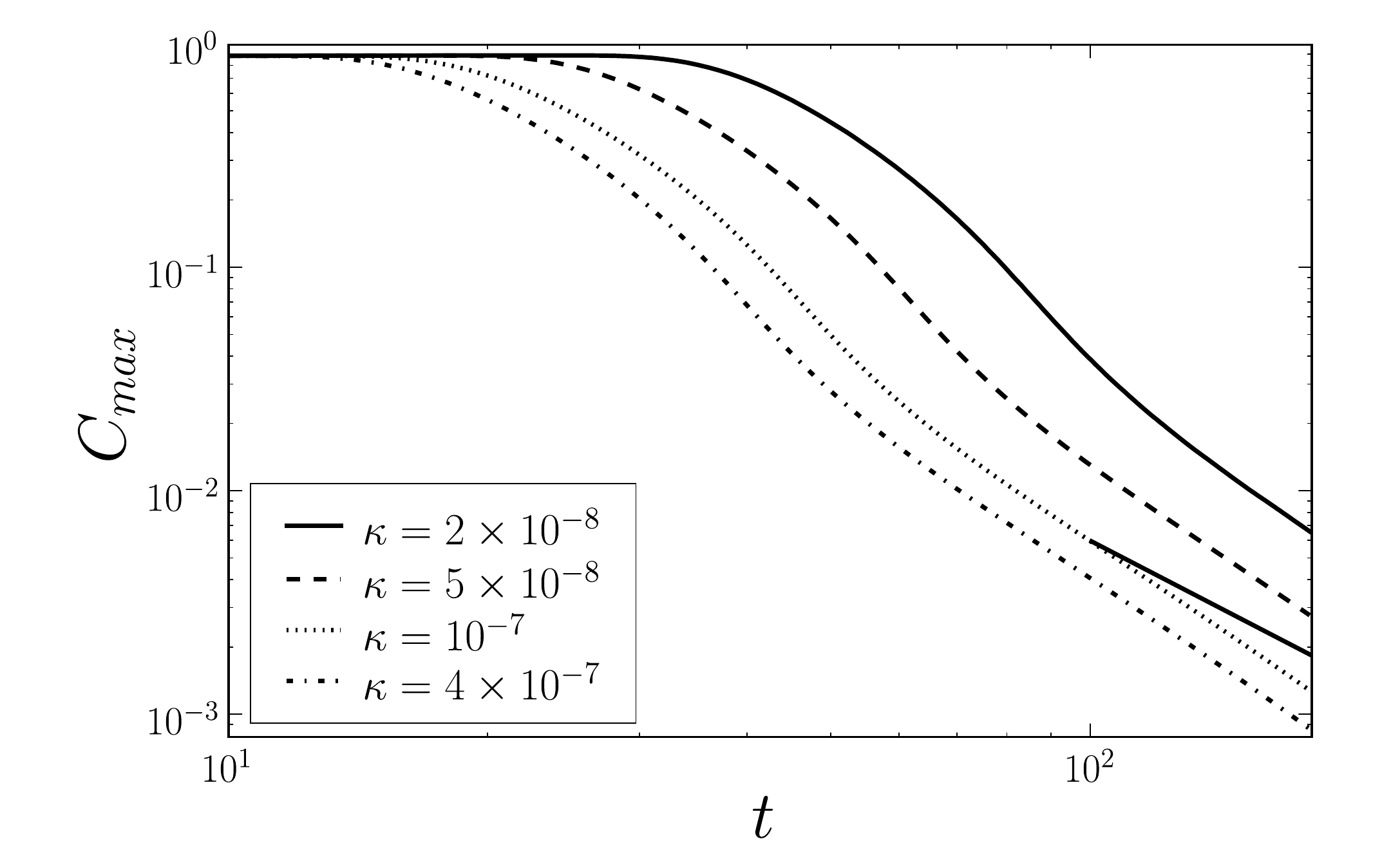}}
}
\caption{Amplitude $C_\mathrm{max}$ of the maximum positive concentration
  fluctuation, in log-linear (a) and log-log (b) coordinates, for different
  diffusivities. This corresponds to the amplitude of the leftmost spike of
  the concentration pattern, which has experienced weaker stretching than
  images of the initial blob in the core of the pattern. An exponential decay
  followed by a power-law evolution are evident. A line of
  slope $-2$ corresponding to the asymptotic evolution expected from Eq.
(\ref{eq:L}) has been drawn
  for comparison. }
\label{figs_closedhom:darkstrip}
\end{figure}

We have plotted in Fig.~\ref{figs_closedhom:darkstrip} the decaying
amplitude of the largest fluctuation in the pattern, that is of the
leftmost box in the mixing pattern
(Figs.~\ref{figs_closedhom:profs}(b) and (c)). The evolution during
200 periods reveals a first exponential decay, whose rate increases
with the diffusivity, followed by a power-law phase with an exponent
of about $-3$. A simple analysis explains this evolution. The
amplitude of this fluctuation can be estimated as $\lac/\wb$, where
$\lac\simeq\prod_{i=1}^t \g_1'(\g_1^i(1))$ is the compression factor experienced after $t$ periods at
the boundary of the pattern (i.e. by the leftmost blob image). (This
is true as long as the distance between the pattern and the wall is
less than the diffusion scale at the boundary. We will discuss this
final phase in Section~\ref{sec:recover}.) For early times dye strips
do not yet feel the effect of the wall, and the stretching factor can
be approximated by $\gamma^t$ (we use $\gamma$ instead of $\Gamma$ for
evaluating the compression by repeated iterations of $\g_1$), and we
expect the decay to be exponential with a rate $\log \gamma$. This
behavior is indeed observed for large enough diffusivities
(Fig.~\ref{figs_closedhom:darkstrip}(a)).  For small diffusivities,
few strips of dye are homogenized before the boundary of the mixing
pattern reaches the wall region, where the effect of the parabolic
point dominates, so we do not observe the first exponential phase.

For long times $\g_1^i(1)\simeq (ai)^{-1}$, $\g_1'(\g_1^i(1))\simeq 1-
2/i$. The compression $\lac$ can be approximated by
\begin{equation}
  \lac \simeq \gamma^{n_0} \prod_{i=n_0+1}^t (1-2/i).
  \label{eq:L}
\end{equation}
The two factors in~\eqref{eq:L} account for (i) the exponential
compression by successive factors of order $\gamma$ inside the bulk,
and (ii) a weaker compression by factors converging slowly to $1$ as
the boundary of the pattern approaches the wall and experiences a
compression determined by the parabolic point at $x=0$
(Section~\ref{sec:nearwall}). The product $\prod_{i=n_0+1}^t (1-2/i)$
converges to a power law $t^{-2}$ for long times. The observed
exponent is greater; this might come from a crossover between an
exponential phase and the predicted $t^{-2}$ phase.

>From the above analysis, we see that positive concentration
fluctuations decay fast with time, compared to the contribution of the
remaining white layer at $x=0$, which shrinks much more slowly. The
contribution of the dark-gray tail to the concentration variance is
very small compared to the light-gray tail, therefore we neglect it in
the following computation of the variance.

It is important to note that the asymmetry of the concentration PDF
persists because our mixer ``remembers'' the initial condition --- a
small black spot and a big white pool --- even after very long times.

In our experiments, additional contributions to the dark-gray tail
come from dye particles trapped during some time in folds of the
pattern where stretching is weak (notice the dark folds in
Fig.\ref{figs_closedhom:homsteps}(a)). This 2-D effect is not present
in our map. The dark-gray tail therefore consists of contributions
from the border of the pattern, but also from these
folds. Nevertheless, we have checked that this contribution is small
compared to the light-gray tail, and decays rapidly with time.

\subsection{Total concentration variance}
\label{sec:alltogether}


We finally sum all contributions from different parts of the PDF to
obtain $\sigma^2(C)$, as in Eq.~\eqref{eq_closedhom:var}. From the
above discussion, we distinguish two phases, $t<\tw$ when the variance
is dominated by the contribution of recently injected white strips
that have not yet reached $\wb$, and $t>\tw$ when the most important
  fluctuations come from the mixing of white strips and gray fluid.

In the experiment, the crossover time $\tw$ is estimated as $30$
periods. However, 3-D effects inside the fluid prevented us from
conducting experiments for more than $35$ periods. For this early
regime, fitting $\sigma^2(C)$ with $\sigma^2_{\mathrm{W}}\propto
(2\log{t}+\log{\wb})/t^2$ (gray line on
Fig.~\ref{figs_closedhom:exp_results}(a)) gives good results, except
close to $\tw$ where the contribution of the light-gray tail starts to
dominate.  On the other hand, in numerical simulations we observe
(Fig.~\ref{figs_closedhom:model_results}(b)) both the
$(2\log{t}+\log{\wb})/t^2$ behavior (black line), which can be
interpreted as in the experiment, and the $t^{-4}$ decay after $\tw$
($100$ periods for the case studied) given by
$\sigma^2_{\mathrm{LG}}$. For long times, the observed power-law
arises from the specific way of incorporating white strips whose width
scales as $t^{-2}$ inside the mixing pattern.

Having established the origin of the scalings observed for the
concentration variance and PDFs in our experiments, we turn in the
next section to the analysis of asymptotically long times, different
initial conditions, optimization, and offer some concluding remarks.


\section{Discussion \label{sec:discuss}}

We have explained in the preceding sections the main features of our
experiment.  In this section we tie two loose ends: we examine the
long-time behavior of the concentration in Section~\ref{sec:recover}, and
look at the effect of initial conditions in Section~\ref{sec:IC}. Both
aspects are more easily investigated in our simple map than in
experiments, and our discussion is supported by numerical simulations of
the 1-D map. In Section~\ref{sec:opt} we address the issue of
optimization of the mixing device based on what we learned about the role
of walls. Finally, we close the paper with concluding remarks in
Section~\ref{disc}.

\subsection{Recovering an eigenmode for long times}
\label{sec:recover}

\begin{figure}
\subfigure[]{
\centerline{\includegraphics[width=8cm]{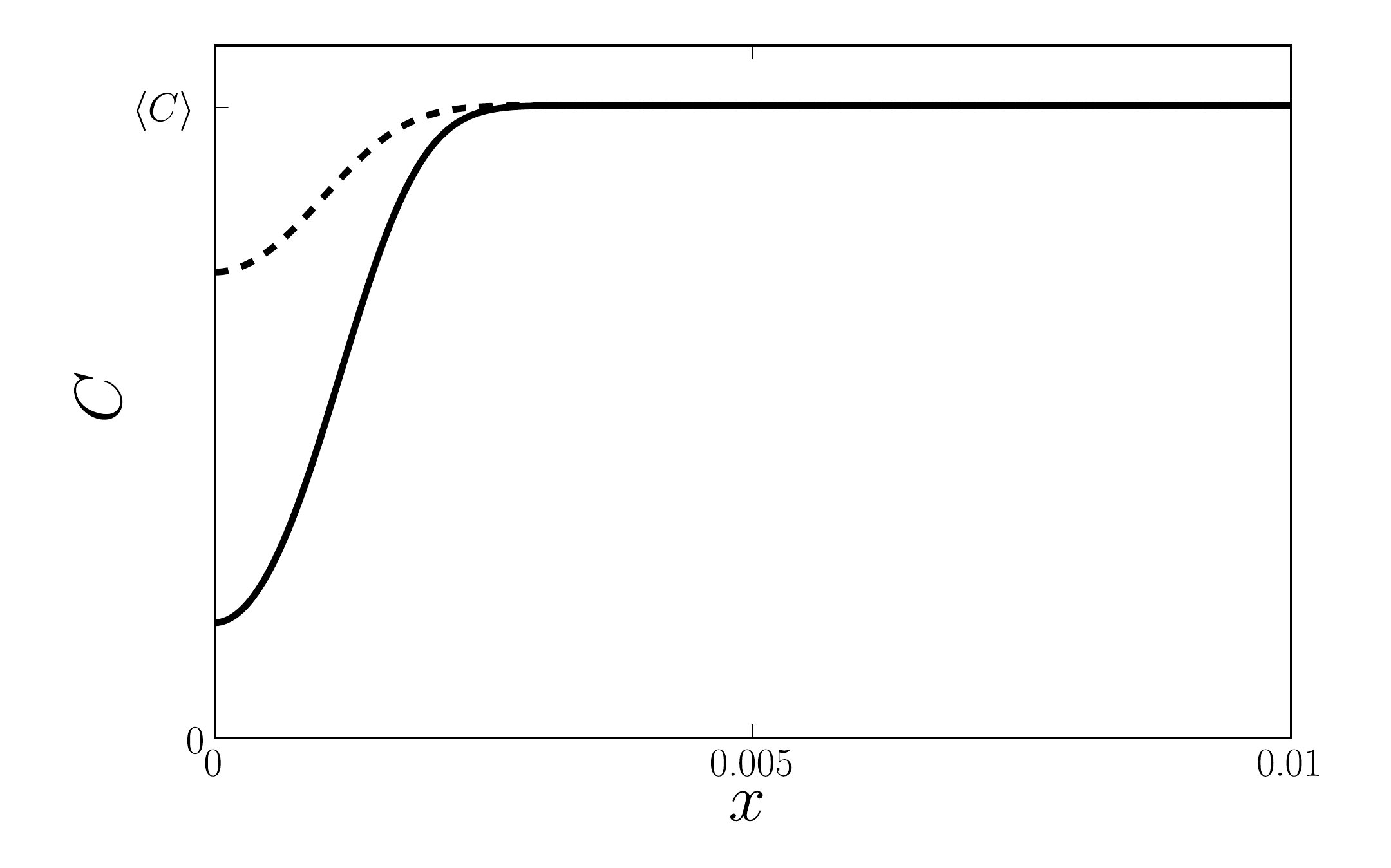}}
}
\subfigure[]{
\centerline{\includegraphics[width=8cm]{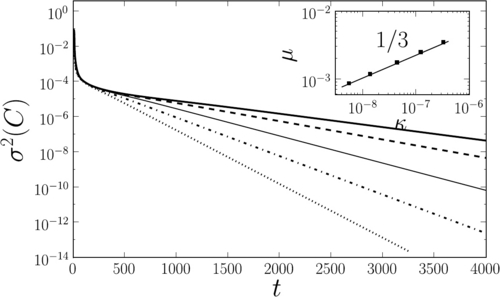}}
}
\caption{(a) Structure of the eigenmode: an inverted
  half-Gaussian of width $w_0$ decays exponentially at a rate
  $-\log(\g_1'(w_0))=2aw_0$. (b) Concentration variance measured in the
  whole unit interval $[0,1]$ for different diffusivities. (The thick
  solid, dashed, solid, dot-dashed and dotted lines correspond
  respectively to the following values of $\kappa$: $5.4\times
  10^{-9}$, $1.3\times 10^{-8}$, $4.3\times 10^{-8}$ $1.2 \times
  10^{-7}$ and $3 \times 10^{-7}$.) For long times, the evolution of
  the variance is exponential, corresponding to the onset of a
  eigenmode: $\sigma^2(C,t)=\sigma^2_0 \exp(-\mueigen t)$. As expected from Eq.~\eqref{eq_closedhom:w0k}, $\mueigen$
  scales as $\kappa^{1/3}$ (inset).
\label{figs_closedhom:eigenplot}}
\end{figure}

We now consider the asymptotic regime, when we can no longer
approximate the injected variance by the contribution of a white strip
of width $1/at^2$.  This is because the mixing pattern is close enough
to the wall that diffusion blurs the white layer at the boundary. For
such large times, the mixing pattern can be described as an inverted
half-Gaussian centered on $x=0$ (see
Fig.~\ref{figs_closedhom:eigenplot}(a)) that decays with time as fluid
is reinjected in the bulk. At this time, fluctuations are very small
in the rest of the pattern, and they are only controlled by the
amplitude of the half-Gaussian. In this final regime, where the
concentration pattern keeps a self-similar form with time, the
concentration profile has eventually converged to an 
eigenmode of the advection-diffusion operator.  The width of the
half-Gaussian $w_0$ is determined by the point where stretching and
diffusion balance,
\begin{equation}
w_0 = \sqrt{\frac{\kappa}{1-\g_1'(w_0)}}.
\end{equation} 
Thus, with $\g_1'(w_0)=1-2aw_0$, we obtain for a small diffusivity
\begin{equation}
w_0 = \l({\kappa}/{4a}\r)^{1/3}.
\label{eq_closedhom:w0k}
\end{equation}
We note that for small diffusivities, $w_0$ -- the Batchelor scale at
$x=0$ -- is much greater than the Batchelor scale in the bulk
$\wb\propto \kappa^{1/2}$.

Once the concentration at $x=0$ starts rising, which occurs for
$d(t)=1/at \sim w_0$ (i.e. $350$ periods for a diffusivity
$\kappa=10^{-7}$!), the stabilized half-Gaussian decays exponentially
at a rate $-\log(\g_1'(w_0))=2aw_0$, 
which scales as
$\kappa^{1/3}$. We have verified this scaling in numerical simulations
of the model (see Fig.~\ref{figs_closedhom:eigenplot}(b)). Note that
this is one of the very few examples where one can predict
analytically the decay of a eigenmode (another noteworthy
situation is the torus map considered in
Ref.~\cite{Thiffeault2003}). However, this eigenmode regime is
not relevant in practice, as we only
observe it when fluctuations are completely negligible in the
bulk. Its structure is also quite trivial: it consists of the
half-Gaussian at $x=0$, and of very small spikes centered on the
iterates of $x=0$ in the bulk.

The convergence to the eigenmode can be interpreted as follows.
Once the mixing pattern reaches the diffusive layer $w_0$ at the
boundary, every box with size equal to the local diffusive scale contains an
iterate of the initial blob of dye. A global decay of the concentration
variance is therefore possible from then on.

\subsection{Other initial conditions}
\label{sec:IC}

For the sake of completeness, we have performed numerical simulations
for other initial conditions than a blob of dye. In particular, we
have simulated two different situations, a cosine profile,
corresponding to an initial condition
\begin{equation}
C(x,t=0)=1+\cos(4\pi x),
\end{equation}
and a random profile where we attribute to each pixel a random value
between 0 and 1. This rapidly varying profile is quickly smoothed
everywhere on the local Batchelor scale.  The main difference between
the two initial conditions may be assessed as follows. For the cosine
profile, the initial scale of variation for the scalar field is much
greater than the Batchelor scale at the boundary $w_0$, whereas for
the random initial condition, the scalar field already varies on the
smallest possible scales.

\begin{figure}
\subfigure[]{
\centerline{\includegraphics[width=8cm]{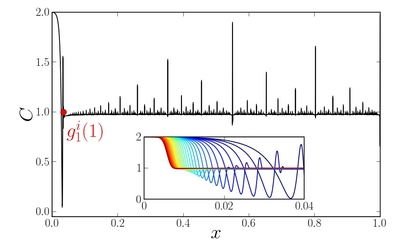}}
}
\subfigure[]{
\centerline{\includegraphics[width=8cm]{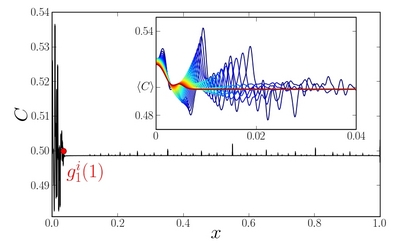}}
}
\caption{[Color online] Homogenization for two different initial conditions: (a)
  $C(x,t=0)=1+\cos(4\pi x)$, (b) random initial profile. The scale of
  variation of the initial profile is much greater than $w_0$ in the
  first case, whereas it is of order $w_0$ close to the boundary in
  the random case. Main axes: concentration profile after 30
  iterations of the map. Note that all variance is contained in the
  leftmost image of the unit interval (always transformed by $\g_1$),
  and around the iterates of $x=1$. Inset: zoom on the border region
  for periods 30 (dark blue) to 180 (red), represented every fifth
  period.\label{figs_closedhom:random_c}}
\end{figure}

In the first case, as for the blob of dye, the scale of variation of
the profile close to the boundary is large, of order $\g^t_1(1)\sim
1/(at)$ (see inset of
Fig.~\ref{figs_closedhom:random_c}(a)). This case is therefore
analogous to the blob of dye case. After a short time, most important
fluctuations are concentrated in the leftmost image of the initial unit
interval, that was iterated only by $\g_1$. Other iterates have
wandered in the bulk where stretching is much more efficient, so that
all fluctuations have died out --- except for newly-reinjected
iterates.  The history of newly-reinjected iterates can be coded as a
sequence $G\circ \g_1 \circ \g_2 \circ \g_1^k$ where $G$ stands for the
last few iterations, which corresponds to the reinjection inside the
bulk of fluctuations at the left boundary. Even the leftmost iterate
feels the spatial heterogeneity of stretching (see inset in
Fig.~\ref{figs_closedhom:random_c}(a)), as fluctuations initially
close to $x=1$ have been more compressed, and they have overlapped and
averaged. After some time, the profile at the boundary (inset in
Fig.~\ref{figs_closedhom:random_c}(a))) has a value significantly
different from the mean concentration only at one or two
``oscillations'', which is exactly what we observe for the blob of dye
case (see Fig.~\ref{figs_closedhom:profs}). As in the latter case, we
observe a power-law decay for the variance evolution, which can be
accounted for by the same reasoning.

For the random profile case, the concentration profile at the boundary
(inset in Fig.~\ref{figs_closedhom:random_c}(b)) is much less coherent
over successive periods. Indeed, the scale of variation of the
concentration profile saturates immediately at the local Batchelor scale
in the whole domain. The strips reinjected in the bulk result from the
averaging of many strips at the boundary, and their amplitude is more
difficult to predict. We measured a non-monotonic decay of variance
inside the bulk in
this case, as the averaging of strips close to the wall depends on the
instantaneous height of many neighboring strips
(Fig.~\ref{figs_closedhom:random_c}(b)). Yet, the strange eigenmode
regime is only reached once fluctuations have died out everywhere, except
for the leftmost box of size $w_0$ and around the iterates of $x=0$, so
that there is a long transient phase also in this case.

However, many features are common to all initial conditions that we
checked. The spatial organization of the bulk profile is dominated by
the unstable manifold of the parabolic point at $x=0$, where
fluctuations persist longer.
As stretching is lower close to the boundary, the
reinjected fluctuations are similar over successive periods. The same
reasoning as for the blob of dye yields concentration PDFs with
power-law tails of the form $|C-\langle C \rangle|^{-2}$, which
are indeed observed in all cases.

\subsection{Hydrodynamical optimization}
\label{sec:opt}

\begin{figure}
\subfigure[]{
\centerline{\includegraphics[width=0.95\columnwidth]{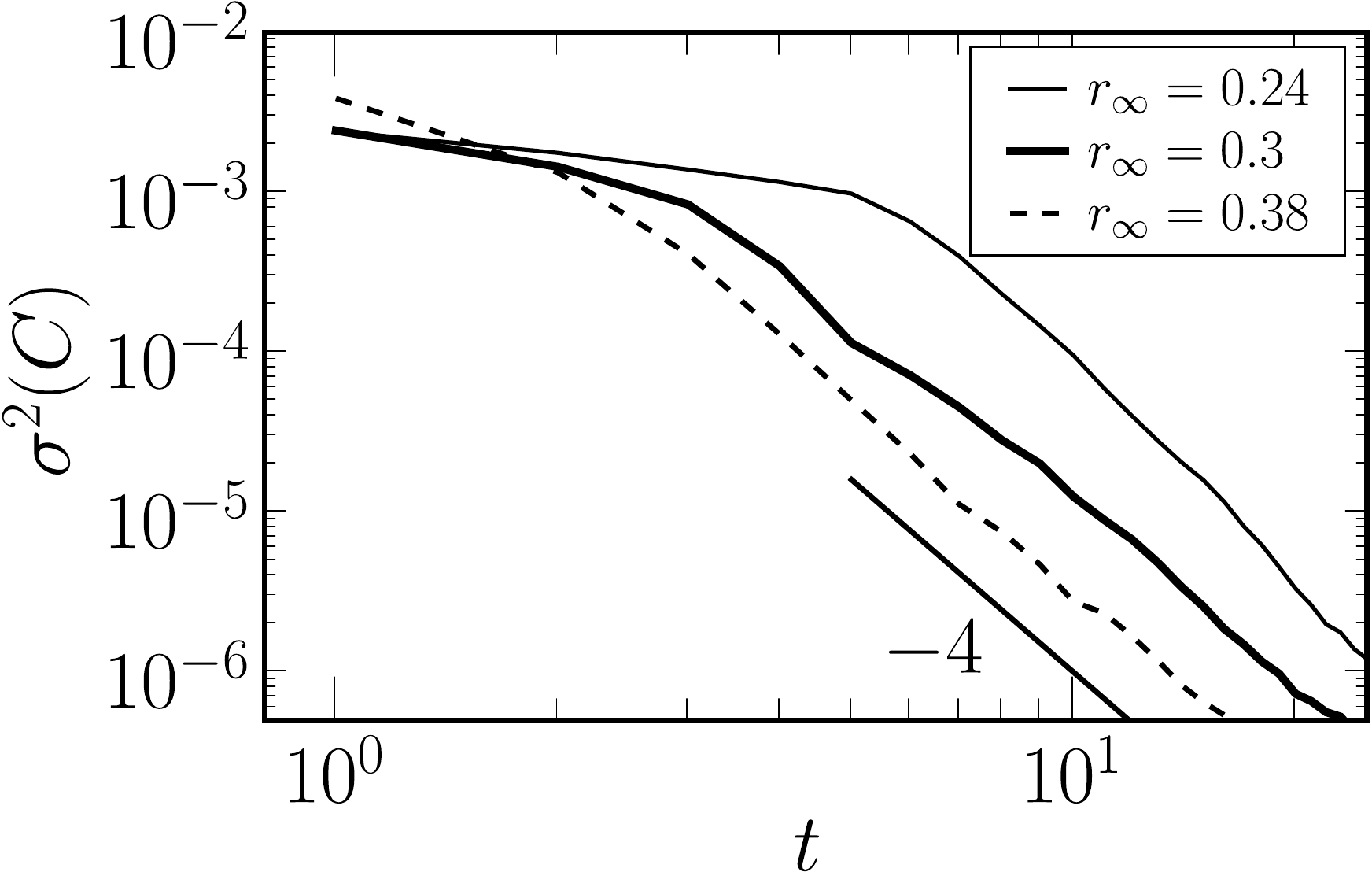}}
}
\subfigure[]{
\centerline{\includegraphics[width=0.95\columnwidth]{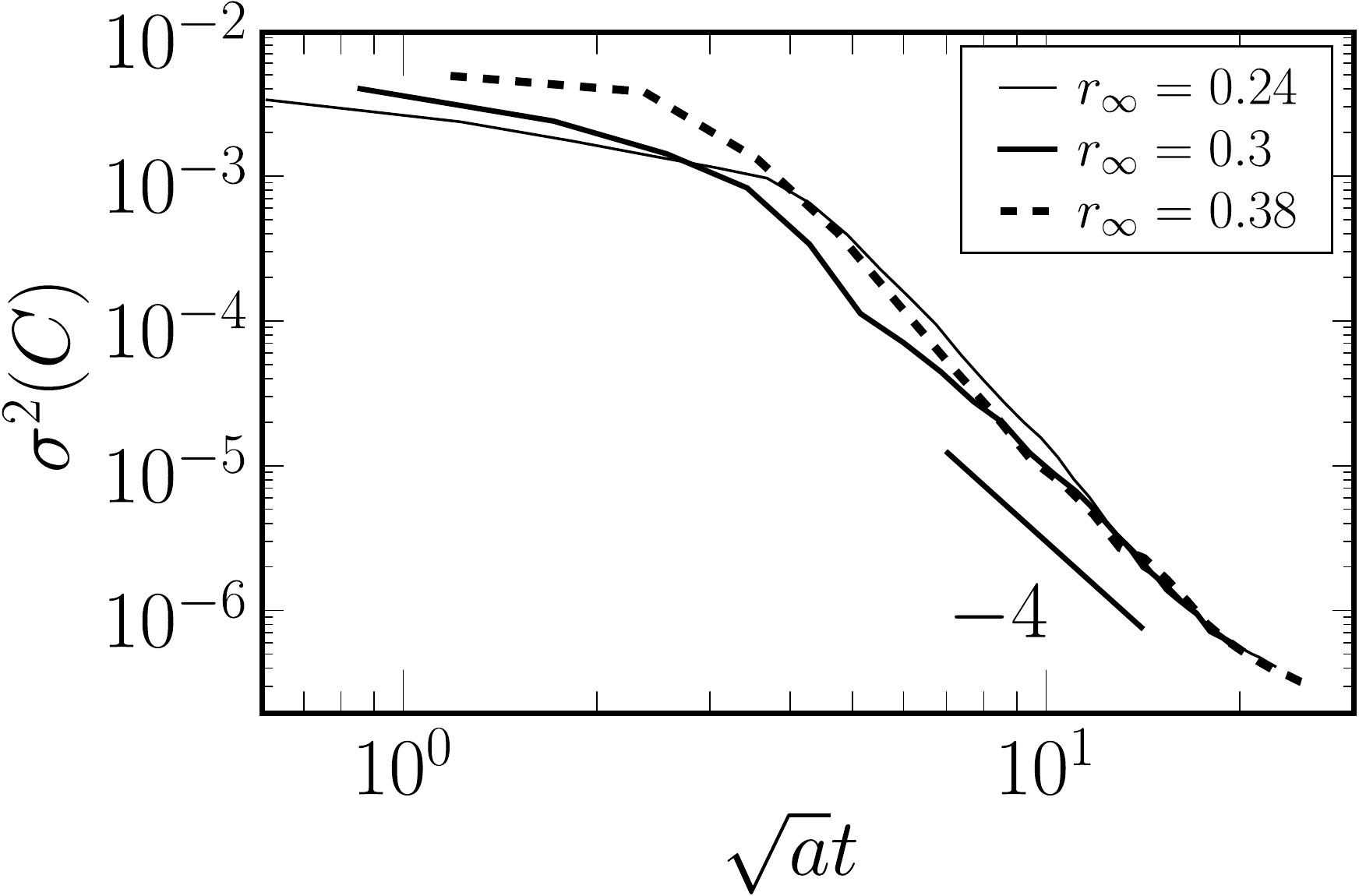}}
}
\caption{(a) Evolution of the concentration variance in a large central
domain for three different versions of the figure-eight protocol,
corresponding to $\reight=0.24,\,0.3$ and $0.38$, obtained by advecting
Lagrangian particles in Stokes-flow numerical
simulations. (b) Evolution of the concentration variance for the same
protocols vs time rescaled by the parameter $\sqrt{\A}$
($\reight=0.24,\,0.3$ and $0.38$ correspond respectively to $\A= 0.39,
0.72$ and $1.39$). A satisfying collapse of all curves
is observed. \label{figs:var_a}}
\end{figure}

We have argued that for a wide class of mixing protocols the decay of
the concentration initially obeys a power law.  For industrial
devices, it is of primary importance to optimize the decay during this
initial phase, that is, to tune the prefactor in the power law. Our
analysis in Section~\ref{sec:Cstat} shows that the prefactor is
essentially determined by the parameter $\A$, which controls the
evolution of the distance between the mixing pattern and the wall in
Eq.~\eqref{eq:d_manip}. (The exponent of the power
law also depends weakly on the mean stretching $\Gamma$.)

We check the validity of the parametrization of the variance by
the rate $\A$ for the figure-eight protocol. As for the blinking
vortex protocol, we follow a large number of particles, in order to
record the evolution of a coarse-grained concentration field. We perform
numerical simulations of a Stokes-flow version of the figure-eight
protocol, for different values of the radius of the figure-eight loops
$\reight$. In all cases, the
initial condition is a small square of size $0.1\times0.1$ located
close to the rod and containing $2.25\times 10^6$ particles.  We
calculate the variance of the concentration on a large half-crown of
outer radius $0.8$.  We show results for the evolution of the variance
in Fig.~\ref{figs:var_a} (a). For $\reight=0.24,\,0.3$ and $0.38$, we
observe a power-law evolution with an exponent close to~$4$, but
slightly greater for the smallest radius $\reight=0.24$. Numerical
simulations do not permit the same spatial resolution as in
experiments; we use a coarse-graining scale $\wb=10^{-2}$. Measuring
the distance of the mixing pattern to the wall, we deduce the values
of $\A$ and $\tw$. Using this coarse spatial resolution and for each
value of $\A$, we compute $\tw\sim 5$. The evolution of the variance
in Fig.~\ref{figs:var_a} is therefore in agreement with the $-4$
exponent of the variance determined in Sec.~\ref{sec:Cstat} for the
regime~$t>\tw$. The larger exponent for the smallest radius could be
attributed to a weaker mean stretching $\Gamma$ in this case, since
the rod travels in a smaller region.

In our model, the dependence of the variance on the
details of the protocol comes from the factor $\stripw^2(t)\simeq
(\sqrt{\A} t)^{-4}$ in the regime $t>\tw$. We have plotted in
Fig.~\ref{figs:var_a} (b) the evolution of the variance against the
rescaled time $\sqrt{\A} t$. We observe a very good collapse of all data on the
same curve, supporting the idea that the main ingredient in the
evolution of the variance is the parameter $\A$.
 
Increasing the mixing speed therefore amounts to increasing the rate
$\A$ at which a particle approaches the wall. This can be achieved in
a number of ways, such as increasing the rod diameter or by ``scraping
the bowl'', that is taking the stirrers closer to the wall as in
Fig.~\ref{figs:var_a}. Determining the value of $\A$ as a function of
hydrodynamic parameters such as the rod diameter or the size of the
rod's orbit is beyond the scope of this article. However, our results
suggest that in comparing different mixing protocols the rate~$\A$
gives a simple estimate of the variance decay rate and can replace
more challenging measurements, such as the concentration field itself.

\subsection{Conclusions}
\label{disc}

The results of this paper can be summarized as follows. Using an
approach based on the Lagrangian description of fluid particles
stretched into filaments, we have highlighted the role of
least-unstable periodic structures in mixing dynamics, first in the
well-known baker's map, and then for a broad class of 2-D closed flows
where the chaotic region extends to a no-slip wall. For the latter
class of systems, we have proposed a generic scenario for
wall-dominated mixing dynamics. No-slip hydrodynamics in the wall
region force poorly-mixed fluid to be slowly reinjected in the bulk along
the unstable manifold of a parabolic point. (Note that phase portraits
with many injection points are also possible. This does not affect
the validity of our arguments.) Mixing dynamics are then controlled by
the slow stretching at the wall, which contaminates the whole mixing
pattern up to its core.  We observed a slow algebraic decay of the concentration
variance in experiments and numerical simulations, which we
justified analytically using a 1-D model of a
baker's map with a parabolic point on the boundary. An exponential
decay corresponding to an eigenmode is recovered in the model
once iterates of the initial blob of dye are present in all boxes of
size of the local diffusive scale, that is for extremely long times at
which the variance has been almost completely exhausted.
   
We characterize a mixing experiment by the following parameters: the flow's mean compression factor
$\Gamma$; the algebraic rate $\A$ at which a particle approaches the boundary;
the Batchelor scale $\wb$ obtained from the diffusivity $\kappa$ and
compression $\Gamma$; and the width of the initial blob $\delta$.
The successive stages of the
mixing process inside the bulk can be summarized as follows.
\begin{itemize}
\item For $t<\log\left(\wb/\delta\right)/\log \Gamma $, 
  all filaments are larger than $\wb$ and the variance is constant:
  $\sigma^2(C)=\sigma^2_0$.
\item For $\log\left(\wb/\delta\right)/\log \Gamma \le t \le \left(\A\wb\right)^{-1/2}$, fluctuations in the bulk start to decay
  as dye filaments are compressed below $\wb$, and the variance is
  dominated by large unmixed strips recently
  injected from the near-wall region into the bulk: $\sigma^2 \propto
  \delta^2\times(\log{\A t^2}+\log{\wb})/(\A\log{\Gamma}\times t^2)$, as
derived in Sec.~\ref{sec:W}.
\item For $\left(\A\wb\right)^{-1/2} \le t \le (\kappa/4\A)^{1/3}$,
  all reinjected strips are smaller than $\wb$, yet their contribution
  still dominates the variance evolution $\sigma^2 \propto 1/(\A^2 t^4)$.
This scaling was derived in Sec~\ref{sec:LG}.
\item For $t\ge (\kappa/4\A)^{1/3} $, we are in the eigenmode regime
(see Sec.~\ref{sec:recover}) and
  $\sigma^2 \propto \exp(-\mueigen t)$, where $\mueigen \propto (\kappa\A^2)^{1/3}$.
\end{itemize}

Our study has highlighted the role of periodic structures
with lowest stretching in the construction of a time-persistent mixing
pattern, dominated by their unstable manifold. This applies to the
least-unstable periodic point in the baker's map (thus a hyperbolic
point) and to the wall parabolic point for the figure-eight case. The
importance of elliptic region for limiting mixing dynamics has been
emphasized in other studies~\cite{Pikovsky2003, Popovych2007}.

In 2-D flows, where Lagrangian dynamics are Hamiltonian, the wall region
can either belong to a chaotic region, or to an elliptical island. We
have argued that algebraic mixing dynamics are obtained in the first
case. An experimental study of mixing dynamics in the second case is in
preparation and will be reported elsewhere.

\subsection*{Acknowledgments}

Natalia Kuncio took an important part in the figure-eight experiments.
The authors also would like to thank C\'ecile Gasquet and Vincent Padilla
for technical assistance, as well as Fran\c{c}ois Daviaud, Emmanuel
Villermaux and Philippe Petitjeans for fruitful discussions. 

\bibliography{journals_abbrev,mixing}
\end{document}